\documentclass[a4paper,fleqn,usenatbib]{mnras}

\usepackage[T1]{fontenc}
\usepackage{ae,aecompl}

\usepackage{graphicx}	% Including figure files
\usepackage{savesym}
\usepackage{amsmath}	% Advanced maths commands
\usepackage{amssymb}	% Extra maths symbols
\savesymbol{iint}
\usepackage{txfonts}
\restoresymbol{TXF}{iint}
\usepackage{bm}

\newcommand{\del}{\partial}
\newcommand{\curl}{\nabla\times}
\newcommand{\subs}{\textsection}
\newcommand{\dv}{\nabla\cdot}
\newcommand{\er}{{\hat{\bf{r}}}}
\newcommand{\ephi}{{\hat{\pmb{\phi}}}}
\newcommand{\ez}{{\hat{\bf{z}}}}
\newcommand{\tomega}{\tilde{\omega}}

\title[]{Quasi-periodic oscillations and the global modes of relativistic, MHD accretion discs}

\author[J. Dewberry et al.]{
Janosz W. Dewberry$^{1}$,\thanks{E-mail: jwd43@cam.ac.uk }
Henrik N. Latter$^{1}$,
Gordon I. Ogilvie$^{1}$\\
\\
$^{1}$DAMTP, University of Cambridge, CMS, Wilberforce Road, Cambridge, CB3 0WA, UK
}

\pubyear{2017}

\begin{document}
\label{firstpage}
\pagerange{\pageref{firstpage}--\pageref{lastpage}}
\maketitle

\begin{abstract} \\
The high-frequency quasi-periodic oscillations (HFQPOs) that punctuate the light curves of X-ray binary systems present a window onto the intrinsic properties of stellar-mass black holes and hence a testbed for general relativity. One explanation for these features is that relativistic distortion of the accretion disc's differential rotation creates a trapping region in which inertial waves (r-modes) might grow to observable amplitudes. Local analyses, however, predict that large-scale magnetic fields push this trapping region to the inner disc edge, where conditions may be unfavorable for r-mode growth. We revisit this problem from a pseudo-Newtonian but fully global perspective, deriving linearized equations describing a relativistic, magnetized accretion flow, and calculating normal modes with and without vertical density stratification. In an unstratified model, the choice of vertical wavenumber determines the extent to which vertical magnetic fields drive the r-modes toward the inner edge. In a global model fully incorporating density stratification, we confirm that this susceptibility to magnetic fields depends on disc thickness. Our calculations suggest that in thin discs, r-modes may remain independent of the inner disc edge for vertical magnetic fields with plasma betas as low as $\beta\approx 100-300$. We posit that the appearance of r-modes in observations may be more determined by a competition between excitation and damping mechanisms near the ISCO than the modification of the trapping region by magnetic fields.\\
\end{abstract}


%%%%%%%%%%%%%%%%%%%%%%%%%%%%%%%%%%%%%%%%%%%%%%%%%%
%%%%%%%%%%%%%%%%% BODY OF PAPER %%%%%%%%%%%%%%%%%%
%%%%%%%%%%%%%%%%%%%%%%%%%%%%%%%%%%%%%%%%%%%%%%%%%%

\section{Introduction}
Understanding the variability observed in the emission from X-ray
binaries remains an important task in astrophysics. So-called
`quasi-periodic oscillations' (QPOs) frequently appear as $0.1$ to
$450$ Hz features in the power density spectrum of stellar-mass black
hole candidate systems. Historically, higher frequency oscillations
(HFQPOs) of $\sim 30-450$Hz, observed in anomalous states of high
accretion and luminosity, have attracted interest for the
comparability of their frequencies to the characteristic orbital frequencies
close to a black hole. Because of their relative insensitivity to luminosity
variations, HFQPOs are thought to issue from
the black hole's imprint on the inner regions of its encircling
accretion disc \citep{ReMCl,Mo16}.

\citet{Oka87} showed that in a purely hydrodynamic and isothermal model strong gravitational effects on the epicyclic frequencies $\kappa$ and $\Omega_z$ create a narrow, annular, `self-trapping' region where large-scale, global inertial waves (sometimes dubbed gravito-inertial modes or g-modes, but here called r-modes) might be constrained to oscillate as standing waves. Protected from dissipation at the inner disc edge by their confinement within this trapping region, hydrodynamic r-modes are subject to excitation by large-scale warping and eccentric deformations in the accretion flow, potentially growing to amplitudes sufficiently large to cause variations in luminosity detectable as HFQPOs \citep{Kat04,Kat08,FoG08,FoG09}. 

However, with temperatures of $\gtrsim1$keV providing sufficient
ionization to support magnetic fields, discs around black holes are
unlikely to be purely hydrodynamic. Magnetohydrodynamic (MHD)
turbulence sustained by the magnetorotational instability (MRI) offers
a widely accepted explanation for an effective viscosity driving
accretion \citep{BH98}. The survival of trapped inertial modes in the
presence of such MHD turbulence is uncertain; in fact, \citet{ReM08} observed
prominent r-mode signatures in hydrodynamic simulations, 
but, along with \citet{Ar06}, found them absent from MRI-turbulent simulations. 
It should be noted, though, that \citet{Ar06} and
\citet{ReM08} omitted any r-mode excitation mechanism other than the
MHD turbulence itself, the latter authors concluding only that the MRI
does  not actively excite the oscillations. On the other hand, 
taking an analytical approach, \citet{FL09} questioned the very
existence of the self-trapping region when the accretion disc is
threaded by large-scale poloidal magnetic fields of appreciable
strength. However, \citet{FL09} presented  
only a local analysis, omitting the radial structure
 of the inherently global oscillations, as well as vertical density stratification. 

In this paper we present a fully global, semi-analytical treatment of
the problem, calculating r-mode oscillations within isothermal
relativistic accretion discs threaded by azimuthal and
vertical magnetic fields. We focus on two models: `cylindrical' discs,
which omit vertical density structure, and fully global discs, which
include it. The wave modes are numerically calculated using pseudo-spectral and hybrid
pseudo-spectral-Galerkin methods, respectively.

In support of the conclusions of \citet{FL09}, we find
that increasingly strong vertical fields force the localisation of r-modes
inwards, toward the ISCO. In cylindrical models, azimuthal fields 
of equipartiton strength have little effect. Meanwhile, r-mode sensitivity 
to the vertical magnetic field depends on the parameterization 
of vertical structure through isothermal sound speed
$c_s$ and vertical wavenumber $k_z$, and is hence somewhat uncertain.
In fully global models, the point at which r-mode confinement
depends on reflection at the inner boundary depends on disc temperature and
thickness. For isothermal sound speeds of $c_s\sim 0.001c$, where
$c$ is the speed of light, we find critical values of the plasma beta 
(ratio of gas to magnetic pressure) as low as
$\beta\sim100-300,$ depending on the scale height's rate of increase
with radius. These estimates are consistent with those of
\citet{FL09}, but we stress that such field strengths are relatively
high for large-scale, ordered magnetic fields; simulations of the MRI
produce fields of this strength or greater, but they are \emph{small-scale}
and \emph{disordered}. Finally, we demonstrate that with a choice of vertical
wavenumber motivated by the basis functions used in our density
stratified analysis, r-modes calculated in a cylindrical model
reproduce the frequencies and localisations of fully global trapped inertial modes to within $1\%$.  

Our results indicate that if the accretion disc is sufficiently thin,
the trapping of r-modes is minimally affected by
large-scale magnetic fields (unless those fields are very strong). And
indeed, QPOs only appear in emission states usually described by thin disc models. The
frequencies, however, of the r-modes will be enhanced by the presence
of large-scale fields, which may complicate their use when divining
the properties of the central black hole. On the other hand, this
frequency enhancement could be used to estimate the strength of the
magnetic field itself, especially if the black-hole mass and spin are known from
other measurements. Finally, we stress that even if r-modes are pushed
up against the ISCO this does not necessarily mean that they are
destroyed. The oscillations will certainly be damped by radial inflow
but could nonetheless achieve appreciable amplitudes if sufficiently
excited by a strong disc eccentricity and/or warp. In short, very strong
large-scale magnetic fields may not exterminate r-modes, 
just make them more difficult to excite. 

The structure of the paper is as follows. In \subs\ref{Bgrnd} we review the nature of global oscillation modes and the potential for their formation in relativistic accretion discs. Readers familiar with the subject are invited to skip to \subs \ref{model}, where we present the linearized equations describing our magnetized, relativistic accretion disc model. In \subs \ref{cylcalc} we examine the effects of azimuthal and vertical magnetic fields on radially global modes calculated in the cylindrical approximation, and in \subs \ref{strcalc} we present our vertically stratified results. Finally, in \subs \ref{conc} we summarize and discuss our findings.

\section{Background}\label{Bgrnd}
Analogous to helioseismology, discoseismology involves the study of accretion disc oscillations that are global in that they maintain their frequency and a coherent structure across a considerable radial extent. Examples might include warps and eccentricities, which can be thought of as non-axisymmetric waves with very low frequency \citep{FoG09}, or large-scale spiral density waves. Subject to sufficient excitation, such oscillations might reach amplitudes large enough to cause observable variations in luminosity. 

The establishment and persistence of global oscillations in a
realistic accretion flow may require specific conditions, however,
especially at the disc boundaries. For
example, the growth of inertial-acoustic oscillations in relativistic
discs via the co-rotation instability requires wave reflection at the
inner radius and the transmission of wave energy at key resonances
\citep{PP,NGG,L13}, while the r-modes considered in this paper may be
damped by radial  inflow at the inner disc edge \citep{FerrThes}.

Accretion discs around stellar-mass black holes offer protection from such damping, the effects of strong gravity possibly providing a narrow region separated from the inner edge within which global r-mode oscillations could reside and grow. While the horizontal epicyclic frequency, $\kappa$, has a monotonic radial profile in Newtonian centrifugally supported discs in near Keplerian rotation, strong gravity introduces radii at which $\kappa^2$ attains a maximum and falls below zero. The latter radius defines the innermost stable circular orbit (ISCO) within which matter plunges toward the black hole, while the existence of the former has spurred the development of several theories of wave propagation aimed at explaining HFQPOs \citep{Ka01}. In this section we review the key elements of these theories, treating hydrodynamic and magnetohydrodynamic models in turn.

\subsection{Hydrodynamic waves in relativistic discs}\label{hydroBg}
\citet{Oka87} used a pseudo-Newtonian treatment with a Paczynski-Wiita potential (see \subs \ref{model}) in considering a hydrodynamic model of a thin, vertically isothermal, relativistic accretion disc. The authors argued that although a rigorous treatment of an accretion disc around a black hole would require a general relativistic model, the effects of strong gravity on wave propagation arise primarily through the relativistic modification of the epicyclic frequency. They showed that the linearized equations derived for perturbations of the form $\delta(r,z)\exp[\textrm{i}m\phi-\textrm{i}\omega t]$, for azimuthal wavenumber $m$ and frequency $\omega$, are approximately separable in $r$ and $z$ under the assumption of a scale height $H(r)=c_s/\Omega_z$ that varies slowly with radius. This separation allows for a description of the perturbations' vertical structure by modified Hermite polynomials of vertical order $n$. Projecting onto a given vertical order leaves a decoupled, second order system of ordinary differential equations in $r$ which, with appropriate boundary conditions, can be solved numerically for the direct calculation of global normal modes.

\begin{figure}                           
        \includegraphics[width=.5\textwidth]{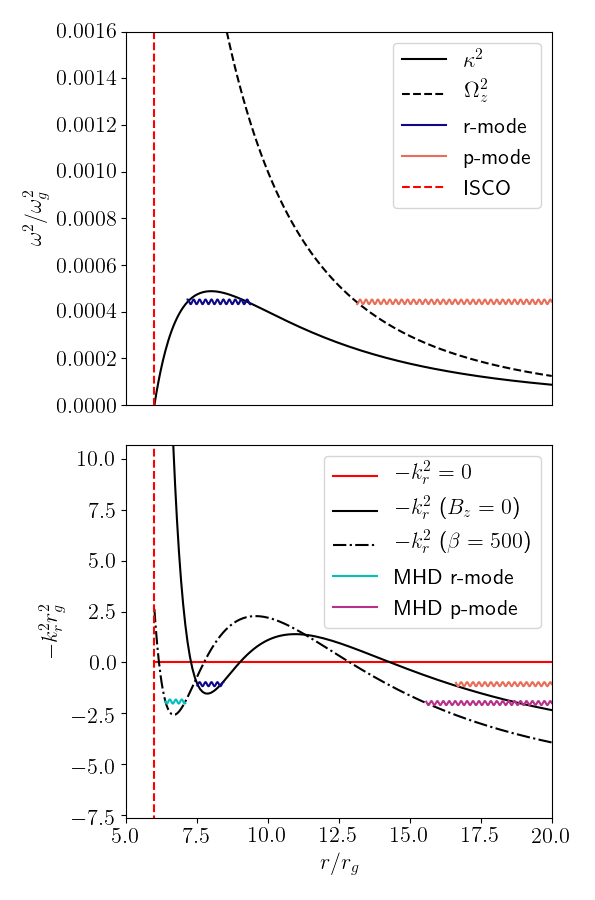}
        \caption{Axisymmetric ($m=0$) wave propagation regions for fully relativistic characteristic frequencies $\kappa_G$ and $\Omega_{Gz}$ (defined in \subs\ref{model}) with spin parameter $a=0$ and sound speed $c_s=0.01c$. R-mode trapping regions, as defined by bounding frequencies $\kappa^2$ and $\Omega_z^2$ (top), and effective potential wells $-k_r^2$ (bottom) are plotted both with no magnetic field (solid lines) and with a vertical magnetic field with midplane $\beta=500$ (bottom, dash-dotted line). Values of $\omega$ taken from the normal mode calculations described in \subs\ref{cylcorr_var} were used to plot the profiles of $-k_r^2$.}\label{WProp}
\end{figure}

Although we are interested in global oscillations, local analyses
provide insight into the effect that the epicyclic frequency's non-monotonic radial profile has on wave propagation. Neglecting the radial variation of background quantities and assuming a radial dependence for perturbations of $\delta(r)\propto\exp[\textrm{i}k_rr]$, where $k_r$ is a local radial wavenumber, \citet{Oka87} derived the dispersion relation
\begin{equation}\label{oDisp}
k_r^2
=\dfrac{\big({\tomega}^2-\kappa^2\big)
\big({\tomega}^2-n\Omega_z^2\big)}
{\tomega^2 c_s^2},
\end{equation}
for $\tomega=\omega-m\Omega$. Since oscillatory behaviour requires real $k_r$, the radial profile for $-k_r^2$ can be thought of as an effective `potential well', allowing wave propagation wherever $-k_r^2<0$ \citep{Li+}. The profiles for the horizontal and vertical epicyclic frequencies $\kappa$ and $\Omega_z$ then define different regions of localisation in the disc for three different families of waves. When $n=0$, 2D `inertial-acoustic' waves can propagate at radii where $\tomega^2>\kappa^2$. Nonzero $n$ gives wave propagation where either $\tomega^2>\max[\kappa^2,n\Omega_z^2]=n\Omega_z^2$ or $\tomega^2<\min[\kappa^2,n\Omega_z^2]=\kappa^2$. The former, higher frequency waves are known as acoustic or p-modes, while the latter, lower frequency waves are the inertial r-modes of our interest. Fig. \ref{WProp} shows the regions of propagation for such global modes in the axisymmetric case, as defined by the conditions on $\omega^2$ (top) and $-k_r^2$ (bottom).

The Lindblad, co-rotation and vertical resonances occur where $\tomega^2=\kappa^2$, $\tomega=0$ and $\tomega^2=n\Omega_z^2$, respectively. They define the wave propagation regions in a hydrodynamic, isothermal disc, and depend closely on the radial profiles of the orbital and epicyclic frequencies. As illustrated in Fig. \ref{WProp}(top), the maximum in $\kappa^2$ (solid line) introduced by the prescription of relativistic frequencies provides a narrow region of confinement for modes with nonzero $n$. The two
Lindblad resonances at radii where $\tomega^2=\kappa^2$ define turning points, within which the inertial modes might be confined irrespective of conditions at the ISCO. The effective potential well defined by $-k_r^2$ for an axisymmetric r-mode with $\omega\approx\max[\kappa]$ (solid line in Fig. \ref{WProp}, bottom) has a minimum at the radius where $\kappa$ achieves its maximum, denoted as $R_{\kappa}.$ This minimum is only local, however, as the vertical resonance occurring at the radius where $\tomega^2=n\Omega_z^2$ allows r-modes to `leak' to the outer regions of the disc and propagate as p-modes. The extent of this leakage is limited by the effective potential barrier between the two propagation regions \citep{FoG08}.

In isolation, trapped inertial waves are almost purely oscillatory, a small exponential decay coming only from wave leakage. Any explanation of HFQPOs involving r-modes then requires an excitation mechanism\footnote{\citet{OrW00} concluded incorrectly (because of a sign error in their analysis) that many hydrodynamic oscillation modes are destabilized by viscosity, but in fact they are damped}. \citet{Kat04,Kat08} estimated the growth rates of trapped inertial waves coupled to large-scale warps or eccentricities in the accretion flow, describing a global analogue to the local parametric instability introduced by \citet{G93} and \citet{RG94}. This coupling involves the transfer of negative wave energy from the fundamental, axisymmetric r-mode, through the global disc deformations, to non-axisymmetric inertial modes propagating within their co-rotation radius. \citet{FoG08} generalized the estimations of \citet{Kat04,Kat08}, calculating explicitly the growth rates of the fundamental trapped inertial mode when excited by warps and eccentricities, and proffering r-modes as a promising explanation for HFQPOs. 

\subsection{Magnetohydrodynamic waves in relativistic discs}\label{mhdBg}
The global calculations conducted by \citet{Oka87} and \citet{FoG08}
bore out the predictions of the local analysis,
bolstering the argument that trapped waves cause
 HFQPOs. However, as recognized by \citet{ReM08} and \citet{FL09}, 
the picture is much more complicated for magnetized accretion flows. 

\citet{ReM08} searched for trapped inertial waves in global
simulations of relativistic accretion discs utilizing a
Paczynski-Wiita potential. Peaks in the power density spectrum at the
frequencies and radii appropriate for r-modes appeared in initially
perturbed, hydrodynamic accretion discs, but were absent from the
spectrum obtained from MHD-turbulent simulations. The authors
concluded that turbulence caused by the MRI does not actively excite
trapped inertial modes, but made no claims about active
damping. Indeed, \citet{OReM} saw trapped r-mode propagation in
viscous simulations with $\alpha>0.05$, but found that the wave
amplitudes were below the noise levels observed in MHD simulations by
\citet{ReM08}. \citet{Ar06} observed a similar absence of inertial
waves in MHD turbulent shearing box simulations. However, like
\citet{ReM08}, these authors did not include any form of excitation
for r-modes other than the MRI turbulence
itself. \citet{H+09} considered such an excitation mechanism in
simulations of MRI turbulent, relativistic discs with nonzero spin,
including a tilt to examine numerically  
the excitation mechanism explored by \citet{FoG08}, and observed
the emergence of modes `at least partially inertial in character.'

Separately, \citet{FL09} put aside the question of MHD
turbulence and argued that large-scale magnetic fields destroy the r-mode
trapping region itself. They derived a local dispersion relation analogous
to Equation (\ref{oDisp}) for a relativistic accretion disc threaded
by a purely vertical, constant magnetic field $B_z$, and found the
self-trapping region modified by an Alfv\'enic restoring force. 
 Rather than being constrained to propagate only where
 $\omega^2<\kappa^2,$ 
their local analyses predict that axisymmetric, 
MHD r-modes are evanescent except in the region where
\begin{equation}\label{bztrap}
k_z^2V_{\textrm{A}z}^2
<\omega^2<
\frac{1}{2}
\left[
    \kappa^2+2k_z^2V_{\textrm{A}z}^2
    +\sqrt{\kappa^4+16k_z^2V_{\textrm{A}z}^2\Omega^2}
\right],  
\end{equation}
where $k_z$ is the vertical wavenumber corresponding to an exponential
dependence $\exp[\textrm{i}k_zz]$, and
$V_{\textrm{A}z}=B_z/\sqrt{\mu_0\rho}$ is the Alfv\'en
speed.\footnote{
  The modes subject to confinement within the
  regions defined by Equation (\ref{bztrap}) are epicyclic-Alfv\'enic
  in nature, but we refer to both them and purely hydrodynamic trapped
  inertial waves as r-modes throughout this paper.} 
The inclusion of a large-scale vertical magnetic field similarly modifies the effective potential well defined by $-k_r^2$, a radial profile of which is plotted with a dash-dotted line in Fig. \ref{WProp}(bottom) for $\beta=500$, the prescription of a vertical wavenumber $k_z=1/H(r)=\Omega_z(r)/c_s$, and a frequency $\omega$ calculated using the modified cylindrical model described in \subs\ref{cylcorr_var}. In the presence of a magnetic field, the potential well extends much closer to the ISCO. For sufficiently strong vertical fields, the trapping region includes the inner edge of the disc, implying that r-mode oscillations would need to be confined by the outer turning point and the inner edge itself, where conditions may be uncertain and most likely unfavourable \citep[e.g.,][]{Ga99,AP03,FerrThes}. 

The self-trapping region's loss of distinction from the inner boundary does not necessarily rule out the existence of r-modes. Modes pushed to the inner edge of the disc would need  to rely on excitation sufficient to overcome the damping effects of radial inflow \citep{FerrThes}, and might require some reflection mechanism for their confinement. Such a reflection is required for explanations of HFQPOs that involve 2D, non-axisymmetric inertial-acoustic modes, which local analyses suggest are less susceptible to the effects of large-scale magnetic fields and subject to amplification at co-rotation \citep{LT09,FL11,L13,YL15}.

In summary, \citet{ReM08} and \citet{FL09} threw the efficacy of
trapped inertial waves as an explanation for HFQPOs into some
doubt. However, the simulations  only indicated that the MRI does not
actively excite r-modes, and in applying local approximations
\citet{FL09} neglected the global nature of discoseismic
oscillations. Local analyses do give insight into the structure of wave
propagation in accretion discs, but the WKBJ ansatz limits us to
scales much smaller
than those of the background flow, an
inappropriate assumption when considering oscillations
coherent across large radial and vertical extents. Additionally,
although the scaling $k_z\propto 1/H$ is natural in the nearly
separable, hydrodynamic problem \citep{Oka87}, the inseparability of
the MHD equations in $r$ and $z$ 
makes the exact proportionality unclear. Further, as noted by
\citet{Kat17}, the inverse dependence of the Alfv\'en speed on density
enhances the importance of vertical density stratification. These
considerations motivate our focus on fully global normal mode
calculations, and our 
 examination of the role that density stratification plays in localising MHD trapped inertial waves.

\section{Disc model and linearized equations}\label{model}
We consider a thin, inviscid, non-self-gravitating, differentially rotating flow. The ideal MHD equations can be written as
\begin{align}
\label{IMHD1}
    \dfrac{\del {\bf{u}}}{\del t}
    +{\bf{u}\cdot\nabla\bf{u}}
    &=-\dfrac{\nabla P}{\rho}
    -\nabla\Phi
    +\dfrac{1}{\mu_0\rho}{\left(\curl\bf{B}\right)\times\bf{B}},
\\[10pt]
    \dfrac{\del\rho}{\del t}
    &=-\dv{(\rho\bf{u})},
\\[10pt]
    \dfrac{\del {\bf{B}}}{\del t}
    &=\curl\left({\bf{u\times B}}\right),
\\[10pt]
\label{IMHD4}
    \dv{\bf{B}}&=0,
\end{align}	
where ${\bf{u}}$, $\rho$, $\bf{B}$, and $P$ are the fluid velocity, density, magnetic field and gas pressure, respectively.

Equations (\ref{IMHD1})-(\ref{IMHD4}) provide a Newtonian description of a plasma in the limit of ideal MHD, and are therefore not strictly applicable to relativistic accretion discs. However, as mentioned in \subs \ref{Bgrnd}, relativistic effects outside of the ISCO are frequently approximated, in both analysis and simulation \citep[e.g.,][]{Oka87,ReM08}, with the prescription of a `pseudo-Newtonian' Paczynski-Wiita potential, given in cylindrical coordinates $(r,\phi,z)$ as
\begin{equation}\label{PW}
    \Phi=\dfrac{-GM}{\sqrt{r^2+z^2}-r_S},
\end{equation}
where $G$ is the gravitational constant, $M$ the mass of the central black hole, and $r_S=2GM/c^2$ is the Schwarzschild radius of the event horizon. In the absence of modification by pressure gradients and magnetic stresses, the introduction of a singularity at $r=r_S$ gives midplane ($z=0$) orbital and horizontal epicyclic frequencies 
\begin{align}
    \Omega_P^2
    &=\dfrac{GM}{r(r-r_S)^2},
\\[10pt]
    \kappa_P^2
    &=\dfrac{GM(r-3r_S)}{r(r-r_S)^3},
\end{align}
the second of which passes through zero at $r=3r_S$, and achieves a
maximum at $r\approx 3.73r_S$.
The point at which $\kappa_P=0$ defines the ISCO, denoted by $r_\text{ISCO}$. The maximum replicates that which appears in the exact frequencies for a particle in orbit around a Kerr black hole, given by
\begin{align}\label{KerrOm1}
    \Omega_G
    &=\dfrac{1}{(r^{3/2}+a)},
\\[10pt]\label{KerrOm2}
    \kappa_G 
    &=\Omega_G\sqrt{1-\dfrac{6}{r}+\dfrac{8a}{r^{3/2}}-\dfrac{3a^2}{r^2}},
\\[10pt]\label{KerrOm3}
    \Omega_{Gz}
    &=\Omega_G\sqrt{1-\dfrac{4a}{r^{3/2}}+\dfrac{3a^2}{r^2}}, 
\end{align}
where $a\in(-1,1)$ is the dimensionless spin parameter, and radii and frequencies are expressed in units of gravitational radius $r_g=GM/c^2$ and the gravitational frequency $\omega_g=c^3/(GM)$. These expressions are commonly inserted in otherwise Newtonian analyses to approximate the effects of black-hole spin on wave propagation in the accretion flow \citep{FoG08}. 

We consider a background equilibrium flow of the form
${\bf{u}}=r\Omega(r)\ephi$, in isorotation with a magnetic field
${\bf{B}}=B_{\phi}(r)\ephi+B_z(r)\ez$. We neglect radial magnetic
fields for their complication of the equilibrium. 
For a gravitational potential of the form given in Equation
(\ref{PW}),
the radial component of Equation (\ref{IMHD1}) then yields 
\begin{equation}\label{OMmmm}
    \Omega^2
    =\dfrac{1}{r}\bigg[
    \dfrac{\del\Phi}{\del r}
    +\dfrac{1}{\rho}\dfrac{\del P}{\del r}
    +\dfrac{1}{\mu_0\rho}\bigg(
    B_{\phi}\dfrac{\textrm{d}B_{\phi}}{\textrm{d}r}
    +B_z\dfrac{\textrm{d}B_z}{\textrm{d}r}
    +\dfrac{B_{\phi}^2}{r}
    \bigg)\bigg].
\end{equation}

For simplicity we assume the gas is globally isothermal with
$P=c_s^2\rho$. In a disc without magnetic fields, a barotropic
equation of state with constant $c_s$ guarantees an equilibrium
angular velocity profile $\Omega$ that depends only on radius. As can
be seen from Equation (\ref{OMmmm}), this remains true when our setup
includes either a density profile $\rho=\rho(r)$ independent of $z$ (as
considered in \subs\ref{cylcalc}), or a purely uniform, vertical
magnetic field (as considered in \subs\ref{strcalc}). In reality, the
temperature, and hence the sound speed, ought to decrease with
radius. We discuss the ramifications of a globally isothermal
equation of state in \subs \ref{conc}; in short, we predict
this assumption leads to an overestimate of the disc scale height as a
function of radius, and consequently an overestimate of r-modes'
susceptibility
to background magnetic fields.

Equation (\ref{OMmmm}) implies that steep gradients in the pressure and magnetic fields, as well as the so-called `hoop stress' due to azimuthal magnetic fields, may cause deviations from the orbits of particles rotating with `pseudo-Keplerian' angular velocities defined by $\Omega_K^2=(1/r)\del_r\Phi$. We include these deviations in calculations utilizing a Paczynski-Wiita potential, but ignore them as $\mathcal{O}(1/r^2)$ effects when utilizing the general relativistic formula for the characteristic frequencies. The vertical independence of ${\bf{u}}$ and ${\bf{B}}$ yields, in the classic thin disc approximation, the hydrostatic equilibrium $c_s^2\del_z \ln\rho=-\del_z \Phi\approx -\Omega_z^2 z$, which gives a background density distribution $\rho(r,z)=\rho_0(r)\exp[-z^2/(2H^2)]$. 

The independence of the equilibrium from $\phi$ and $t$ allows us to
consider perturbations of the form
$\delta(r,\phi,z,t)=\delta(r,z)\exp[\text{i}m\phi-\text{i}\omega
t]$. Because non-axisymmetric r-modes encounter a co-rotation
resonance and may be strongly damped at the radius where
$\tomega=\omega-m\Omega=0$ \citep{Li+}, historical focus has been
given to axisymmetric trapped inertial waves with $m=0$. The
co-rotation resonance might be avoided by non-axisymmetric modes with
frequencies such that $\omega=m\Omega(r)$ lies outside of the
self-trapping region, but such frequencies would be on the order of
kHz, too high to explain HFQPO observations in stellar-mass black-hole
binaries. For this reason, we restrict our attentions to axisymmetric
modes. In particular, we focus on the fundamental axisymmetric r-mode
with the simplest radial and vertical structure,
as it is the least likely to be disrupted, and the most likely to produce a net luminosity variation. 

We disturb the equilibrium with axisymmetric fluid velocity, magnetic
field and enthalpy ($h\equiv\delta P/\rho$) perturbations of the form
$[{\bf{v}},{\bf{b}},h](r,z)\exp[-\textrm{i}\omega t]$,
 where $\omega$ is in general complex. Linearising Equations
(\ref{IMHD1})-(\ref{IMHD4}) then yields
\begin{align}
\label{lEqns1}
    &-\textrm{i}\omega v_r
    =2\Omega v_{\phi}
    -\left[
        \dfrac{\del }{\del r}
        -\dfrac{1}{\mu_0P}
        \left(B_z\dfrac{\textrm{d}B_z}{\textrm{d}r}+\dfrac{B_{\phi}}{r}\dfrac{\textrm{d}(rB_{\phi})}{\textrm{d}r}\right)
    \right]h
\\
\nonumber
    &\qquad
    +\dfrac{1}{\mu_0\rho}
    \Bigg[
        B_z\dfrac{\del b_r}{\del z}
        -\left(B_{\phi}
            \dfrac{\del }{\del r}
            +\dfrac{\textrm{d}B_{\phi}}{\textrm{d}r}+\dfrac{2B_{\phi}}{r}
        \right)b_{\phi}
        -\left(
            B_z\dfrac{\del }{\del r}+\dfrac{\textrm{d}B_z}{\textrm{d}r}
        \right)b_z
    \Bigg], 
\\[10pt]
\label{lEqns2}
    &-\textrm{i}\omega v_{\phi}
    =-\dfrac{\kappa^2}{2\Omega}v_r
    +\dfrac{1}{\mu_0\rho}
    \left(
        \dfrac{1}{r}\dfrac{\textrm{d}(rB_{\phi})}{\textrm{d} r}b_r
        +B_z\dfrac{\del b_{\phi}}{\del z}
    \right),
\\[10pt]
    &-\textrm{i}\omega v_z
    =-\dfrac{\del h}{\del z}
    +\dfrac{1}{\mu_0\rho}
    \left(
        \dfrac{\textrm{d}B_z}{\textrm{d} r}b_r
        -B_{\phi}\dfrac{\del b_{\phi}}{\del z}
    \right),
\end{align}
\begin{align}
    &-\textrm{i}\omega h
    =-c_s^2
    \left\{
        \left[
            \dfrac{\del}{\del r}+\dfrac{1}{r}
            \left(
                1+\dfrac{\textrm{d} \ln\rho_0}{\textrm{d} \ln r}
            \right)
        \right] v_r
        +\left(
            \dfrac{\del }{\del z}
            +\dfrac{\del \ln\rho}{\del z}
        \right)v_z
    \right\},
\\[10pt]
    &-\textrm{i}\omega b_r
    =B_z\dfrac{\del v_r}{\del z},
\\[10pt]\label{lEqns6}
    &-\textrm{i}\omega b_{\phi}
    =-\left(
        B_{\phi}\dfrac{\del}{\del r}
        +\dfrac{\textrm{d}B_{\phi}}{\textrm{d}r}
    \right)v_r
    +B_z\dfrac{\del v_{\phi}}{\del z}
    -B_{\phi}\dfrac{\del v_z}{\del z}
    +\dfrac{\textrm{d}\Omega}{\textrm{d}\ln r}b_r,
\\[10pt]\label{lEqns7}
    &-\textrm{i}\omega b_z
    =-\left(
        B_z\dfrac{\del}{\del r}
        +\dfrac{\textrm{d}B_z}{\textrm{d}r}
        +\dfrac{B_z}{r}
    \right)v_r,
\end{align}
where $\dv{\bf{b}}=0$ has been substituted into the induction equation, so that the perturbations satisfy the solenoidal condition by construction. In the following sections, we solve Equations (\ref{lEqns1})-(\ref{lEqns7}) under various approximations and in full.

\section{Cylindrical calculations}\label{cylcalc}
To set the scene and isolate clearly the radially global aspects of
MHD r-modes, we first present calculations made in the cylindrical
approximation, in which the vertical lengthscale of the perturbations
is assumed to be much smaller than that of the equilibrium
flow. Strictly speaking, this assumption is inappropriate; the r-modes
of most interest are those with the simplest vertical structure. (These are the least affected by large-scale magnetic fields.) However, the cylindrical model is particularly attractive from a numerical standpoint, providing a less expensive framework for global, non-linear simulations. Furthermore, we show in \subs\ref{cyl_corr} and \subs\ref{cylcorr_var} that with a specific choice of  vertical wavenumber, cylindrical r-modes can be very closely identified with trapped inertial waves calculated 
from a fully global and self-consistent model.

Neglecting the vertical dependence of $\Phi$ and $\rho$ in Equations
(\ref{lEqns1})-(\ref{lEqns7}), we 
Fourier transform and prescribe a plane-wave $z$-dependence $\exp[\textrm{i}k_zz]$, where the vertical wave number $k_z$ is assumed for simplicity and self-consistency to be constant (revisited in \subs\ref{cylcorr_var}). We do consider radial variation in ${\bf{B}}$ and $\rho$ in making calculations with a Paczynski-Wiita potential, writing
\begin{align}
    {\bf{B}}
    &=B_{\phi 0}\left(\dfrac{r}{r_0}\right)^q\ephi
    +B_{z0}\left(\dfrac{r}{r_0}\right)^p\ez,
\\[10pt]
    \rho
    &=\rho_0\left(\dfrac{r}{r_0}\right)^{\sigma}, 
\end{align}   
where $r_0$ is the radius of the ISCO in the absence of modification by gas pressure and magnetic forces, $B_{\phi0}$ and $B_{z0}$ are constant magnetic field strengths, and $q,$ $p$ and $\sigma$ are constant power law indices. For convenience, we follow \citet{FL09} in defining the midplane Alfv\'enic Mach number $\mathcal{M}_A=|{\bf{V_{\textrm{A}}}}|/c_s=\sqrt{2}\beta^{-1/2}$, where ${\bf{V_A}}={\bf{B}}/\sqrt{\mu_0\rho}$ is the midplane Alfv\'en velocity and $\beta$ is the midplane plasma beta. 
We then write $\mathcal{M}_{\textrm{Az}}(r)=V_{\textrm{A}z}/c_s$ and $\mathcal{M}_{\textrm{A}\phi}(r)=V_{\textrm{A}\phi}/c_s$, noting that both are functions of radius. With these prescriptions, the equilibrium flow subject to a Paczynski-Wiita potential follows the midplane angular velocity 
\begin{equation}\label{cylOm}
\Omega^2=\dfrac{GM}{r(r-r_S)^2}
+\dfrac{c_s^2}{r^2}\big[\sigma
+\big(q+1\big)\mathcal{M}_{\textrm{A}\phi}^2+p\mathcal{M}_{\textrm{A}z}^2\big].
\end{equation}
Note that we only use this expression when $p,q,\sigma\neq 0$. When
omitting all background radial gradients in $\bf{B}$ and $\rho$ we employ the correct general relativistic frequencies, setting $\Omega\approx\Omega_G$ and $\kappa\approx\kappa_G$, from Equations \eqref{KerrOm1}-\eqref{KerrOm3}.

Trading the components of ${\bf{b}}$ for the Alfv\'en velocity perturbation ${\bf{v}}_A={\bf{b}}/\sqrt{\mu_0\rho(r)}$, Equations (\ref{lEqns1})-(\ref{lEqns7}) can be written as
\begin{align}
    \notag
    -\textrm{i}\omega v_r
    &=2\Omega v_{\phi}
    -\left(
        \dfrac{\textrm{d} }{\textrm{d}r}
        -\dfrac{(q+1)}{r}\mathcal{M}_{\textrm{A}\phi}^2-\dfrac{p}{r}\mathcal{M}_{\textrm{Az}}^2
    \right)h
\\\notag
    &\hskip.5cm
    +c_s
    \Bigg[
        \textrm{i}k_z\mathcal{M}_{\textrm{Az}}v_{\textrm{A}r}
    -\mathcal{M}_{\textrm{A}\phi}
        \left(
            \dfrac{\textrm{d} }{\textrm{d} r}+\dfrac{\sigma+2q+4}{2r}
        \right)v_{\textrm{A}\phi}
\\\label{lCylvBeqns1}
    &\hskip3cm
    -\mathcal{M}_{\textrm{Az}}
        \left(
            \dfrac{\textrm{d} }{\textrm{d} r}+\dfrac{\sigma+2p}{2r}
        \right)v_{\textrm{A}z}
    \Bigg],
\\[10pt]
    -\textrm{i}\omega v_{\phi}
    &=-\dfrac{\kappa^2}{2\Omega}v_r
    +c_s
    \left(
        \dfrac{(q+1)}{r}\mathcal{M}_{\textrm{A}\phi}v_{\textrm{A}r}
        +\textrm{i}k_z\mathcal{M}_{\textrm{Az}}v_{\textrm{A}\phi}
    \right),
\\[10pt]
    -\textrm{i}\omega v_z
    &=-\textrm{i}k_zh
    +c_s
    \left(
        \dfrac{p}{r}\mathcal{M}_{\textrm{Az}}v_{\textrm{A}r}
        -\textrm{i}k_z\mathcal{M}_{\textrm{A}\phi}v_{\textrm{A}\phi}
    \right),
\\[10pt]
    -\textrm{i}\omega h
    &=-c_s^2
    \left[
        \left(
            \dfrac{\textrm{d}}{\textrm{d} r}+\dfrac{1+\sigma}{r}
        \right) v_r
        +\textrm{i}k_zv_z
    \right],
\\[10pt]
    -\textrm{i}\omega v_{\textrm{A}r}
    &=\textrm{i}k_zc_s
    \mathcal{M}_{\textrm{Az}}v_r,
\\[10pt]
    -\textrm{i}\omega v_{\textrm{A}\phi}
    &=-c_s
    \left[ 
        \mathcal{M}_{\textrm{A}\phi}
        \left( 
            \dfrac{\textrm{d}}{\textrm{d}r}+\dfrac{q}{r}
        \right)v_r
        -\textrm{i}k_z\mathcal{M}_{\textrm{Az}}v_{\phi}
        +\textrm{i}k_z\mathcal{M}_{\textrm{A}\phi}v_z
    \right]
    \nonumber 
    \\&\hskip5cm
    +\dfrac{\textrm{d}\Omega}{\textrm{d}\ln r}v_{\textrm{A}r},
\\
\label{lCylvBeqns7}
    -\textrm{i}\omega v_{\textrm{A}z}
    &=-c_s\mathcal{M}_{\textrm{Az}}
    \left(
        \dfrac{\textrm{d}}{\textrm{d} r}+\dfrac{p+1}{r}
    \right)v_r.
\end{align}
Non-dimensionalizing radii, frequencies, and velocities in units of $r_g,$ $\omega_g$, and $r_g\omega_g=c$ (resp.), Equations (\ref{lCylvBeqns1})-(\ref{lCylvBeqns7}) can be written as a generalized eigenvalue problem of the form ${\bf{A}\cdot{U}}=\omega {\bf{B\cdot U}}$, where ${\bf{B}}$ specifies the boundary conditions placed on the eigenvector ${\bf{U}}$, and solved using a pseudo-spectral method. With derivatives approximated by Chebyshev spectral derivative matrices, this requires only the use of standard numerical packages for finding the eigenvalues and eigenvectors of matrices \citep{Boyd}. All normal mode calculations made in the cylindrical approximation were performed using a Gauss-Lobatto grid on $r\in[r_{\textrm{ISCO}},r_{\textrm{ISCO}}+14r_g]$ with a discretization of $N=200$.

\subsection{Cylindrical, hydrodynamic r-modes}
We point out that the hydrodynamic dispersion relation in the cylindrical approximation is similar to Equation (\ref{oDisp}), but with $n\Omega_z^2$ replaced by $k_z^2c_s^2.$ In our simplified model with constant $k_z$, the vertical resonance has vanished: $k_z^2c_s^2$ does not decrease with radius like $n\Omega_z^2$, so p-modes with $\omega^2>k_z^2c_s^2>\kappa^2$ can propagate freely, while r-modes with $\omega^2<\kappa^2$ are evanescent everywhere except in the narrow region near the radius of maximal $\kappa,$ $R_{\kappa}$. As a result, the r-modes calculated with the cylindrical model experience no decay from wave leakage, and $\omega$ is entirely real. This justifies the use of the rigid wall boundary condition $v_r=0$ at the outer as well as the inner boundaries in hydrodynamic calculations using the cylindrical approximation. 

Apart from the absence of wave leakage, our hydrodynamic trapped
inertial modes resemble those of \citet{FoG08}. The solid curves in
Fig. \ref{zz0_modes}(top,middle) show example radial profiles of $Re[v_r]$ for
normal modes. Here the calculations employ the correct general
relativistic formulas for the frequencies $\Omega_G$ and $\kappa_G$. Though hydrodynamic calculations are not the focus of this paper, we briefly summarize the effects of different parameters as a point of comparison with previous work. Increasing $k_z$ causes the fundamental trapped inertial mode's confinement to narrow, and its frequency to grow. Increasing $c_s$, which causes the potential well $-k_r^2$ to become wider, expands the radial extent of the r-modes. Using a Paczynski-Wiita potential to investigate radial density variation, we find that in the hydrodynamic model, physically reasonable power law indices $0>\sigma\geq-1.5$ give a pressure gradient that acts only minimally through $\Omega$ and $\kappa$.

\subsection{Cylindrical, magnetohydrodynamic r-modes}
\subsubsection{Magnetohydrodynamic boundary conditions}\label{MHDbc}
Although nonzero magnetic fields introduce more derivatives to Equations (\ref{lCylvBeqns1})-(\ref{lCylvBeqns7}), with appropriate choices of variable the system can be reduced to a single ODE of second order \citep[e.g.,][]{ApVa,OgP96}. We then require exactly two boundary conditions, and considering a local dispersion relation for a radial wavenumber $k_r$  (derived from Equations (\ref{lCylvBeqns1})-(\ref{lCylvBeqns7}) with the assumptions $\delta(r)\propto\exp[\textrm{i}k_rr]$, $k_r,k_z\gg1/r$) provides insight into which may be appropriate \citep[for the case with $B_{\phi}=0$, see Equation 28 in ][]{FL09}.
\footnote{
We note that the profiles for $-k_r^2$ found from this dispersion relation can deviate based on whether or not the classic equality $d\Omega/d\ln r=\kappa^2/(2\Omega)-2\Omega$ is used in conjunction with the general relativistic versions of the characteristic frequencies. Although this equality was assumed in deriving Equations (\ref{lEqns1})-(\ref{lEqns7}), it is not accurate for characteristic frequencies $\kappa_G,$ $\Omega_G$ derived from the Kerr metric (Equations \ref{KerrOm1}-\ref{KerrOm3}). In utilizing our dispersion relation and solving Equations (\ref{lEqns1})-(\ref{lEqns7}), we instead equate the direct formulae for $\kappa_G^2$ and $d\Omega_G/d\ln r$ with the epicyclic and shearing terms appearing in Equations (\ref{lEqns2}) and (\ref{lEqns6}) (resp.).
} 

For frequencies within the range defined by Equation (\ref{bztrap}) and constant  $k_z\gtrsim\Omega_z(r_0)/c_s$, we find that effective potential wells defined by $-k_r^2$ remain positive at $r_{\textrm{out}}=r_{\textrm{ISCO}}+14r_g$ for the magnetic field strengths considered. This is because our assumption of a constant $k_z$ still excludes a vertical resonance. The choice of outer boundary condition then makes little difference. Sufficiently strong vertical magnetic fields can drive the profiles for $-k_r^2$ to negative values at the ISCO, however, suggesting that MHD r-mode confinement may become dependent on reflection at the inner boundary. To assess the degree of this dependence, we also consider the boundary conditions $db_r/dr=0$ and $dv_{\phi}/dr=0$ at $r_{\textrm{in}}=r_{\textrm{ISCO}}$, utilizing the former unless otherwise stated.

\begin{figure}                          
    \includegraphics[width=.5\textwidth]{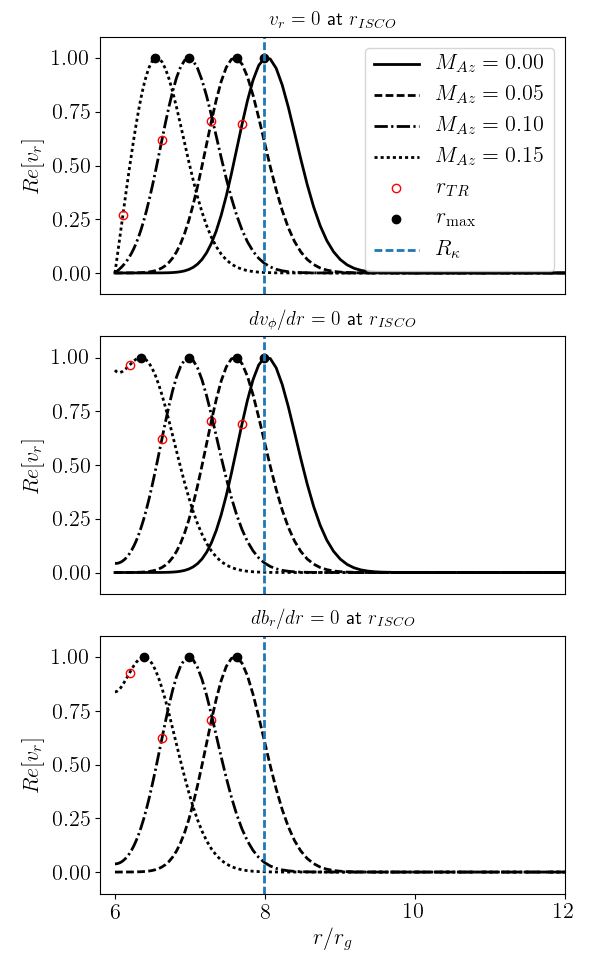}
    \caption{Radial profiles of $Re[v_r]$ for hydrodynamic and magnetohydrodynamic r-modes calculated with nonzero $B_z$, with inner and outer boundary conditions given as $v_r=0$ (top), $dv_{\phi}/dr=0$ (middle), and $db_r/dr=0$ (bottom). General relativistic characteristic frequencies were used ($\Longrightarrow p,q,\sigma=0$), with $a=0$ and  $k_z=\Omega_z(r_0)/c_s\approx 22.7r_g^{-1}$ for $c_s=0.003c$. The filled black and unfilled red dots give the radii of maximal $v_r$ (denoted $r_{\textrm{max}}$) and radius where $-k_r^2=0$ for the calculated frequency $\omega$ (denoted $r_{\textrm{TR}}$).}\label{zz0_modes}
\end{figure}

\subsubsection{Purely vertical magnetic field, $B_{\phi}(r)=0,B_z(r)\not=0$}\label{BzCyl}

We use the same pseudo-spectral method as earlier to calculate
solutions with a nonzero vertical magnetic field. As in the
hydrodynamical case, each calculation at a given $k_z$ produces a discrete set of modes, each with radial structures of increasing complexity. The modes may be
distinguished by quantum numbers $l$, where $l=0$ corresponds to the
fundamental, which exhibits the simplest structure.

Fig. \ref{zz0_modes} shows plots of $Re[v_r]$ for fundamental magnetohydrodynamic r-modes calculated with a purely constant vertical magnetic field of strengths up to $\mathcal{M}_{\textrm{Az}}=0.15$ ($\beta\lesssim100$), using general relativistic formulas for the characteristic frequencies, all three boundary conditions, $c_s=0.003c$ and the naive assumption
$k_z=\Omega_z(r_0)/c_s\approx 22.7r_g^{-1}$. The frequencies
$Re[\omega]$ can be used to
construct profiles for $-k_r^2$, which give a local estimate, here
called $r_{\textrm{TR}}$, of the r-mode turning point, or inner edge
of the trapping region (plotted as red, unfilled dots in
Fig. \ref{zz0_modes}). The black dots give the modes' radius of
maximal radial velocity, $r_{\max}$. While $r_{\max}$ does not measure
the evanescence of the r-modes, unlike $r_{\textrm{TR}}$ it provides 
 a global measure of their localisation, and qualitatively illustrates the effect of the vertical magnetic field.

\begin{figure}                          
    \includegraphics[width=.5\textwidth]{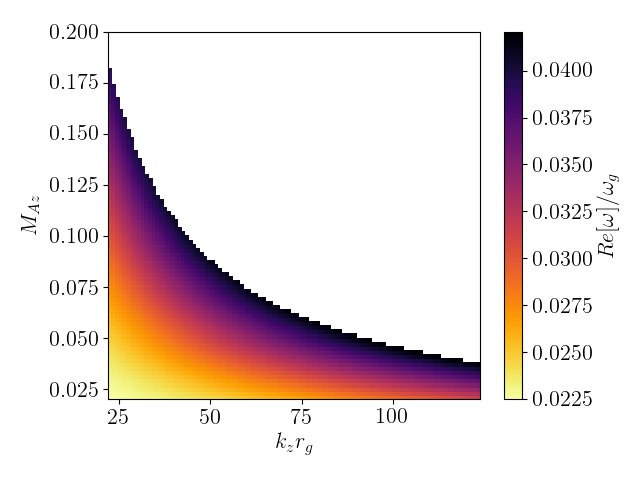}
    \caption{$Re[\omega]$ for the fundamental $l=0$ r-mode calculated in the cylindrical approximation, as in Fig. \ref{zz0_modes}(bottom),
   for varying $k_z$ and Alfv\'enic Mach number  $\mathcal{M}_{\textrm{Az}}$. Modes found to have $r_{\textrm{max}}\leq r_{\textrm{ISCO}}$ are  excluded.}\label{zz0_kz_Omz}
\end{figure}

Fig. \ref{zz0_modes} indicates that the inclusion of a purely constant
vertical magnetic field forces r-modes toward the plunging region
within the ISCO, as predicted by the local analysis. This migration is accompanied by changes in mode frequency, illustrated in Fig. \ref{zz0_kz_Omz}. These alterations to the mode behaviour occur because the addition of a vertical magnetic field amplifies the restoring force to horizontal disturbances, causing the r-modes to become hybrid epicyclic-Alfv\'enic, and their frequencies to increase with $B_z$ as well as $k_z.$ If the Alfv\'enic restoring force is too large, the modes cease to be confined without reflection at the inner boundary (compare the middle and bottom panels of Fig. \ref{zz0_modes} with the top, although a nonzero value of $v_r$ at the inner boundary does not necessarily mean the mode is not evanescent there). We find that the frequency increases shown in Fig. \ref{zz0_kz_Omz} follow the dependence on Alfv\'en frequency $\omega_{\textrm{A}z}=k_zV_{\textrm{A}z}$ given by Equation (\ref{bztrap}), but diverge by as much as $10\%$ for larger values of $k_z$ and $V_{Az}.$

\begin{figure}                          
    \includegraphics[width=.5\textwidth]{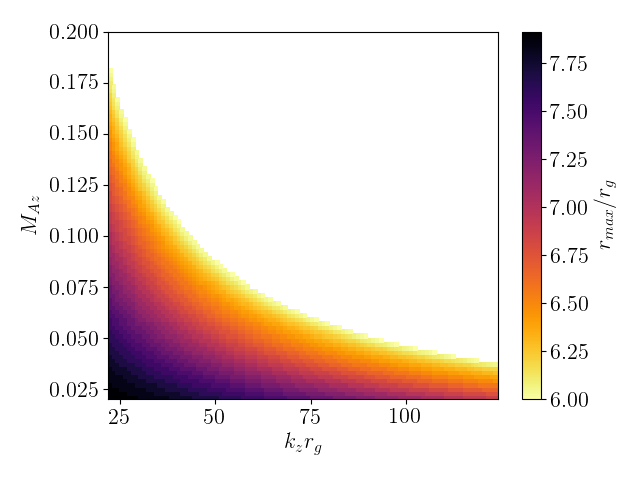}
    \caption{Radius of maximal $v_r$, $r_{\max }$, for the fundamental
      $l=0$ r-mode calculated in the cylindrical approximation, as in Fig. \ref{zz0_modes}(bottom), for varying $k_z$ and Alfv\'enic Mach number  $\mathcal{M}_{\textrm{Az}}$. Modes found to have $r_{\textrm{max}}\leq r_{\textrm{ISCO}}$ are  excluded.}\label{zz0_kz_Urmax}
\end{figure}

The degree to which the mode's localisation is affected by $\mathcal{M}_{\textrm{Az}}$ depends on both the vertical wavenumber $k_z$ and the sound speed $c_s$, the combination of which parameterizes vertical structure in the cylindrical model. Fig. \ref{zz0_kz_Urmax} shows a heatmap of $r_{\textrm{max}}$ for the modes whose frequencies are shown in Fig. \ref{zz0_kz_Omz}. The two figures show a similar dependence of the mode frequency and localisation on $\mathcal{M}_{Az}$ and $k_z$. The sound speed chosen has a more significant impact on $r_{\textrm{max}}$ than on $\omega$, however. Naively choosing $k_z=\Omega_z(r_0)/c_s$ and varying $c_s$ and $\mathcal{M}_{\textrm{Az}}$, we find that at a given $k_z$, lower values of $c_s$ make the r-modes marginally less susceptible to the vertical magnetic field, as measured by $r_{\max}$. More importantly, larger values of $c_s$ increase the width of the trapping regions for the r-modes, causing them to encounter the inner boundary at lower magnetic field strengths. Including a nonzero spin parameter also prolongs the r-modes' march towards the ISCO.

Performing calculations with a Paczynski-Wiita potential yields qualitatively the same results, and allows for the convenient evaluation of the effects of radial density and magnetic field variation, which modify both the rotation profile (\ref{cylOm}) and Equations (\ref{lCylvBeqns1})-(\ref{lCylvBeqns7}). Radial density variation with power laws $0\geq\sigma\geq-1.5$ impact MHD r-modes more than in the hydrodynamic case, acting through the Alfv\'en speed to delay the point at which $r_{\textrm{max}}\sim r_{\textrm{ISCO}}$, and to marginally increase their frequencies further. Radially decreasing $B_z$ acts also through the background Alfv\'en velocity $V_{\textrm{A}z}$, but instead forces the fundamental r-mode closer to the ISCO at a given value of $B_{z0}$. This can be understood through Equation (\ref{bztrap}). As the r-modes migrate inward, the restoring force from the magnetic field becomes larger and larger.

\subsubsection{Purely toroidal magnetic field, $B_{\phi}(r)\not=0,B_z(r)=0$}
Trapped inertial mode calculations made with $B_{\phi}\not=0,B_z=0$ and rigid wall boundary conditions
confirm the predictions of \citet{FL09} that large-scale azimuthal
magnetic fields modify r-modes only negligibly. For the ranges of sound speed and vertical wavenumber considered in \subs \ref{BzCyl}, an equipartition field strength with $\beta=1$ ($\mathcal{M}_{A\phi}=\sqrt{2}$) is required for even a $0.05\%$ change in frequency. Changes in r-mode localisation are imperceptible.

\section{Vertically stratified, fully global calculations}\label{strcalc}
Our calculations of trapped inertial waves made in the ideal MHD,
cylindrical approximation confirm the qualitative behaviour predicted
by the local analyses of \citet{FL09}. Although our simple cylindrical
model provides an inexpensive framework for non-linear simulations, it
nevertheless has too many free parameters, and is strictly
inappropriate for the case of most interest with $k_z\sim1/H$. To find
accurate measurements of the critical field strengths at which the
mode confinement becomes dependent on the inner boundary, then, it is
essential to consider the vertical structure of the disc. In this
section we present a fully general, vertically stratified model for
axisymmetric oscillations in an accretion disc threaded by a uniform
vertical magnetic field in \subs \ref{stEqns} and \subs\ref{expBCs},
and calculations of critical magnetic field strengths 
both without and with the coupling of vertical modes provided by radial scale height variation in \subs \ref{cstH} and \subs \ref{varH}  (resp.).

In this section, we also seek to validate the use of an unstratified,
cylindrical model to study the behaviour of trapped inertial waves and
other global oscillations. The ansatz of an exponential dependence
$\delta(r,z)\propto\tilde{\delta}(r)\exp[\textrm{i}k_zz]$ relies on
the assumption that the background flow's scale of vertical variation
is much larger than that of the perturbations. Cylindrical models may
then be understood to target phenomena taking place at the midplane of
the disc. For this approach to be valid, we must find a correspondence
between r-modes calculated with both the cylindrical model, and a
fully general treatment. In particular, we must identify an
appropriate vertical wavenumber. The choice of $k_z$ greatly
influences cylindrical r-modes' response to increasing vertical
magnetic field strength (see \subs \ref{cylcalc}), but, as discussed
in \subs \ref{Bgrnd}, is made unclear by the non-separability
introduced by such a field. In \subs \ref{cyl_corr} and \subs
\ref{cylcorr_var}, we present calculations demonstrating a clear
correspondence both without and with (resp.) the inclusion of radial scale height variation.

The importance of vertical density stratification has been recognized by \citet{Kat17}, who noted that the Alfv\'en speed associated with a purely constant vertical magnetic field would have a much steeper vertical than horizontal gradient in an isothermal disc. \citet{Kat17} also considered a vertically stratified, isothermal accretion disc threaded by a purely constant vertical magnetic field, but used radially local analyses, and asymptotic expansions with $\mathcal{M}_{\textrm{Az}}=V_{\textrm{A}z}/c_s$ as a small parameter. Further, the author used the argument of a rarified corona to impose a rigid lid. Here, we instead present fully global calculations made without explicit restrictions on the magnetic field strength, and relax this imposition of vertical disc truncation.

\subsection{Stratified equations: a change of variables}\label{stEqns}

Since our cylindrical analyses suggest that both azimuthal magnetic fields and radial variation in $\rho$ and $\bf{B}$ have very little effect on the trapped inertial modes, we consider a purely constant ${\bf{B}}=B_z\ez$ and set $p=q=\sigma=0$. Thus the density profile is radially constant: $\rho=\rho_0g(\eta)$, where $\eta(r,z)\equiv z/H(r)$ is a new vertical coordinate, and $g=\exp[-\eta^2/2]$ is the Gaussian density profile associated with an isothermal disc.

In addition to this change of coordinate, we modify Equations
(\ref{lEqns1})-(\ref{lEqns7}) with a change of dependent variable. In
the case of a purely constant ${\bf{B}}=B_z\ez,$ the perturbation to
the Lorentz force ${\bf{J}}\times\bf{B}$ is proportional to $(\curl{\bf{b}})\times{\bf{B}}$, prompting the definition
\begin{equation}
{\bf{L}}
\equiv 
B_z\left[
    \dfrac{1}{H}\dfrac{\del b_r}{\del \eta}
    -\left(
        \dfrac{\del }{\del r}
        +\dfrac{\del \eta}{\del r}\dfrac{\del }{\del \eta}
    \right) b_z
\right]\er 
+\dfrac{B_z}{H}\dfrac{\del b_{\phi}}{\del \eta}\ephi.
\end{equation}
Eliminating $b_{\phi}$ and $b_z$ in favour of ${\bf{L}}$'s $r$ and $\phi$ components, $L_r$ and $L_{\phi}$, and trading the enthalpy perturbation $h=c_s^2\delta\rho/\rho$ for the non-dimensional $\Gamma=\delta\rho/\rho$ for convenience, Equations (\ref{lEqns1})-(\ref{lEqns7}) can be re-written as 
\begin{align}
\label{LorEqnsH1}
    -\textrm{i}\omega v_r
    &=2\Omega v_{\phi}
    -c_s^2\left(
        \dfrac{\del }{\del r}
        -\dfrac{\textrm{d}\ln H}{\textrm{d} r}\eta\dfrac{\del }{\del \eta}
    \right)\Gamma 
    +\dfrac{L_r}{\mu_0\rho},
\\[10pt]
    -\textrm{i}\omega v_{\phi}
    &=-\dfrac{\kappa^2}{2\Omega}v_r
    +\dfrac{L_{\phi}}{\mu_0\rho},
\\[10pt]
    -\textrm{i}\omega v_z
    &=-\dfrac{c_s^2}{H}\dfrac{\del \Gamma}{\del \eta},
\\[10pt]
    -\textrm{i}\omega \Gamma 
    &=-\mathcal{L}_rv_r
    +\dfrac{\textrm{d} \ln H }{\textrm{d} r}\dfrac{\eta}{g}\dfrac{\del(gv_r)}{\del \eta }
    -\dfrac{1}{gH}\dfrac{\del (gv_z)}{\del \eta },
\\[10pt]
    -\textrm{i}\omega b_r
    &=\dfrac{B_z}{H}\dfrac{\del v_r}{\del \eta},
\\[10pt]
    -\textrm{i}\omega L_r
    &=B_z^2\left\{
        \mathcal{L}_{rr}
        -\mathcal{L}_{HH}\eta\dfrac{\del }{\del \eta}
        +\left[
            \dfrac{1}{H^2}
            +\left(\dfrac{\textrm{d}\ln H}{\textrm{d}r}\right)^2\eta^2
        \right]\dfrac{\del^2}{\del \eta^2}
    \right\}v_r,
\\[10pt]\label{LorEqnsH7}
    -\textrm{i}\omega L_{\phi}
    &=\dfrac{B_z^2}{H^2}\dfrac{\del^2 v_{\phi}}{\del \eta^2}
    +\dfrac{B_z}{H}\dfrac{\textrm{d}\Omega}{\textrm{d}\ln r}\dfrac{\del b_r}{\del \eta},
\end{align}
where we have defined the differential operators 
\begin{align}
    \mathcal{L}_r
    &\equiv\dfrac{\del }{\del r}+\dfrac{1}{r},
\\[10pt]
    \mathcal{L}_{rr}
    &\equiv\dfrac{\del^2 }{\del r^2}
    +\dfrac{1}{r}\dfrac{\del }{\del r}
    -\dfrac{1}{r^2},
\\[10pt]
    \mathcal{L}_{HH}
    &\equiv\dfrac{\textrm{d}\ln H}{\textrm{d}r}
    \left(2\dfrac{\del }{\del r}
    +\dfrac{1}{r}
    -2\dfrac{\textrm{d}\ln H}{\textrm{d}r}
    \right)
    +\dfrac{1}{H}\dfrac{\textrm{d}^2H}{\textrm{d}r^2}.
\end{align}

\subsection{Boundary conditions and series expansions}\label{expBCs}
In this section we describe the series expansions and numerical method utilized in solving Equations (\ref{LorEqnsH1})-(\ref{LorEqnsH7}). Readers interested only in the results of our calculations may skip to \subs\ref{cstH} and \subs\ref{varH}. For numerical efficiency we follow \citet{Oka87} and \citet{OgBe} in making the ansatz that for a given perturbation $\delta(r,\eta),$ we can write
\begin{equation}
\delta(r,\eta)
=\sum_n^{\infty}\delta_n(r)\mathcal{B}_n(\eta),    
\end{equation}
where $\{\mathcal{B}_n\}$ is a set of orthogonal basis functions
satisfying the boundary conditions imposed on $\delta(r,\eta)$ as
$\eta\rightarrow\pm\infty.$ Finding the exact solution would require
solving for an infinite number of radially variable coefficients
$\delta_n(r),$ but we find that with the appropriate choice of
$\mathcal{B}_n$, truncating the series expansions at a finite $n=M$
can produce excellent approximations to the eigenvalues $\omega$. 

In the hydrodynamic case, \citet{Oka87} showed that the fluid
variables ${\bf{v}}$ and $h$ are well described by modified Hermite
polynomials $\textrm{He}_{n}$ of order $n$. However, in the
magnetohydrodynamic problem, the $r$ and $\phi$ components of the
induction equation imply that velocities increasing polynomially with
$\eta$ would result in an unbounded bending of the field lines high
above the disc. Although bending may in reality occur for large-scale
poloidal field loops with footpoints in the outer disc, our focus on
the inner radii of the black hole accretion disc prompts us to impose
the condition that the magnetic field lines become purely vertical at
large $\eta.$ This implies that
$b_r,b_{\phi},\del_zv_r,\del_zv_{\phi}\rightarrow 0$ as
$|\eta|\to\infty$. 
A more appropriate choice, then, is to expand $v_r,v_{\phi}$ and $\Gamma$ as 
\begin{align}
    v_r(r,\eta)
    &=\sum_{n=1}^{\infty}u_n(r)F_n(\eta),
\\[10pt]
    v_{\phi}(r,\eta)
    &=\sum_{n=1}^{\infty}v_n(r)F_n(\eta),
\\[10pt]
    \Gamma(r,\eta)
    &=\sum_{n=1}^{\infty}\Gamma_n(r)F_n(\eta),
\end{align}
where $u_n$, $v_n$, and $\Gamma_n$ are functions to be determined. 
The basis functions $F_n(\eta)$ are solutions to the equation 
\begin{equation}
\dfrac{\textrm{d}^2F}{\textrm{d}\eta^2}+K^2gF=0,   
\end{equation}
where $K$ is a constant, dimensionless eigenvalue. Subject to the boundary condition that $\del_{\eta}F\rightarrow0$ as $\eta\rightarrow\infty$ this equation \citep[derived by][]{GB94,Og98,Latter10} is in Sturm-Liouville form, with guaranteed discrete sets $\{F_n\}$ and $\{K_n\}$ \citep[written as \{$K_nH$\} in][]{Latter10}. The $F_n$ describe the horizontal velocity components of MRI channel modes in the local, anelastic approximation, and are related to another set of basis functions, $\{G_n(\eta)\},$ by 
\begin{equation}
\dfrac{\textrm{d}F_n}{\textrm{d}\eta}=K_nG_n.
\end{equation}
Both sets of functions (cf. Fig. \ref{FnGn} and fig. $1$ in \citet{Latter10}) are orthogonal, and can be normalized such that 
\begin{align}
\label{orth1}
    \int_{-\infty}^{\infty}gF_nF_md\eta&=\delta_{nm},
\\[10pt]\label{orth2}
    \int_{-\infty}^{\infty}G_nG_md\eta&=\delta_{nm}.
\end{align}
The functions $\{G_n\}$ provide an appropriate basis set for the radial magnetic field perturbation, which we expand as
\begin{equation}
b_r(r,\eta)=\sum_{n=1}^{\infty}b_r^n(r)G_n(\eta).
\end{equation}

For the Lorentz force perturbation, we appeal to the models of \citet{GB94} and \citet{SanoM}, who assumed that far from the disc the magnetic field goes to a force free configuration, such that ${\bf{L}}\rightarrow0$. Finally, for the vertical velocity perturbation, we note that the assumption that $\Gamma$ goes to a constant implies that $v_z\rightarrow0$ as $\eta\rightarrow\pm \infty$ as well. In theory, then, $v_z,$ $L_r$ and $L_{\phi}$ can be expanded in any set of orthogonal basis functions that go to zero at infinity, but we find the fastest convergence for
\begin{align}
    v_z(r,\eta)&=\sum_{n=1}^{\infty}w_n(r)g^{1/4}(\eta)F_{n-1}(\eta),
\\[10pt]
    L_{r}(r,\eta)&=\sum_{n=1}^{\infty}L_{r}^n(r)g(\eta)F_n(\eta),
\\[10pt]
    L_{\phi}(r,\eta)&=\sum_{n=1}^{\infty}L_{\phi}^n(r)g(\eta)F_n(\eta),
\end{align}
where $w_n$, $L_r^n$, and $L_\phi^n$ are functions to be
determined. 

It should be noted that $G_0=0,K_0=0,$ and $F_0$ is a constant. Since $F_{-1}$ is undefined, the expansion for $v_z$ must start from $n=1.$ For the vertically structured inertial waves, the $n=0$ basis functions do not describe any of the other perturbations, and their exclusion has no impact on the solutions. 

The basis functions for $v_z$ and ${\bf{L}}$ 
are also orthogonal with the weights $g^{1/2}$ and $g^{-1}$ (resp.). Substituting our expansions into Equations (\ref{LorEqnsH1})-(\ref{LorEqnsH7}), we gain an infinite number of coupled systems of seven equations, but this orthogonality allows us to greatly reduce the dimensionality of our problem. First scaling $r$ and $H$ by $r_g$, frequencies by $\omega_g$, ${\bf{v}}$ and $c_s$ by $r_g\omega_g=c,$ $b_r$ by $B_z$, and ${\bf{L}}$ by $B_z^2/r_g$, we multiply Equations (\ref{LorEqnsH1})-(\ref{LorEqnsH7}) by suitable combinations of basis functions of arbitrary order $m$ and integrate over $\int_{-\infty}^{\infty}d\eta$ to gain (after swapping indices)
\begin{align}
\label{LorEqnsHexp1}
    -\textrm{i}\omega u_n
    &=2\Omega v_n
    -c_s^2\dfrac{\textrm{d}\Gamma_n }{\textrm{d} r}
    +c_s^2\dfrac{\textrm{d}\ln H}{\textrm{d} r}\sum_{m=1}^{\infty}\nu_{nm}\Gamma_m
    +\mathcal{M}_{\textrm{Az}}^2c_s^2L_r^n,
\\[10pt]
    -\textrm{i}\omega v_n
    &=-\dfrac{\kappa^2}{2\Omega}u_n
    +\mathcal{M}_{\textrm{Az}}^2c_s^2L_{\phi}^n,
\\[10pt]
    -\textrm{i}\omega w_n
    &=-\dfrac{c_s^2}{H}\sum_{m=1}^{\infty}\theta_{nm}\Gamma_m,
\\[10pt]
    -\textrm{i}\omega \Gamma_n
    &=-\mathcal{L}_ru_n
    -\dfrac{\textrm{d}\ln H}{\textrm{d} r}\sum_{m=1}^{\infty}\lambda_{nm}u_m
    +\dfrac{1}{H}\sum_{m=1}^{\infty}\alpha_{nm}w_m,
\\[10pt]
    -\textrm{i}\omega b_r^n
    &=\dfrac{K_n}{H}u_n,
\\[10pt]
    -\textrm{i}\omega \sum_{m=1}^{\infty}\dfrac{\gamma_{nm}}{K_m^2}L_r^m
    &=\mathcal{L}_{rr}u_n \notag\\
&    -\sum_{m=1}^{\infty}\left[
        \nu_{nm}\mathcal{L}_{HH}
        +\dfrac{1}{H^2}\gamma_{nm}
        +\left(\dfrac{\textrm{d}\ln H}{\textrm{d}r}\right)^2\mu_{nm}
    \right]u_m,
\\[10pt]\label{LorEqnsHexp7}
    -\textrm{i}\omega L_{\phi}^n
    &=-\dfrac{K_n^2}{H^2}v_n
    -\dfrac{K_n}{H}\dfrac{\textrm{d}\Omega}{\textrm{d}\ln r}b_r^n,
\end{align}
where the coefficients $\theta_{nm},$ $\alpha_{nm},$ $\gamma_{nm},$
$\lambda_{nm},$ $\nu_{nm}$ and $\mu_{nm}$ are defined by finite
integrals involving various combinations of $\eta$, $g$, $F_n,F_m,G_n$
and $G_m$ (see Appendix \ref{Acouple}). These coupling coefficients
connect the equation sets of different orders, and thus illustrate the
non-separability of the problem, though as we shall see the coupling
is relatively weak. A given set of of equations of $n$'th order couples to other sets of equations with orders $n+2,n+4,...$ of the same parity, but the integrals quickly go to zero away from $n\approx m$. This suggests that we may lose little in accuracy by truncating the series in Equations (\ref{LorEqnsHexp1})-(\ref{LorEqnsHexp7}) at a relatively small $m=M$. A danger with this numerical method is that inappropriately chosen basis functions can lead to un-converged solutions. In general, however, we find that with these basis expansions, calculating a converged trapped inertial mode of a given vertical order $n$ only requires extending the series expansion to $M\geq n+4$. With truncation at larger $M$, energetic contributions to the solution from coefficients of higher order are negligible, as are changes in frequency (further discussion in Appendix \ref{converge}). 

Discretizing our radial domain on a Gauss-Lobatto grid with $N$ gridpoints, Equations (\ref{LorEqnsHexp1})-(\ref{LorEqnsHexp7}) can once again be written as a generalized eigenvalue problem  ${\bf{A}\cdot\bf{U}}=\omega{\bf{B}\cdot\bf{U}}$, where $\bf{B}$ now carries the coupling coefficients $\gamma_{nm}$ in addition to encoding the boundary conditions, of which we require two for each order $n$ included in the truncated series. The solution takes the form ${\bf{U}}
=({\bf{U}}_1,{\bf{U}}_2,...,{\bf{U}}_M)^T$ with  $n$'th order radial coefficients 
${\bf{U}}_n
=({\bf{u}}_n,{\bf{v}}_n,{\bf{w}}_n,{\bf{h}}_n,{\bf{b}}_r^n,{\bf{L}}_r^n,{\bf{L}}_{\phi}^n
)^T$. 
We find that $10$ grid points per $r_g$ is sufficient to resolve the least radially complicated, fundamental r-mode of physical interest.

\begin{figure*}
    \centering
    \includegraphics[width=.99\textwidth]{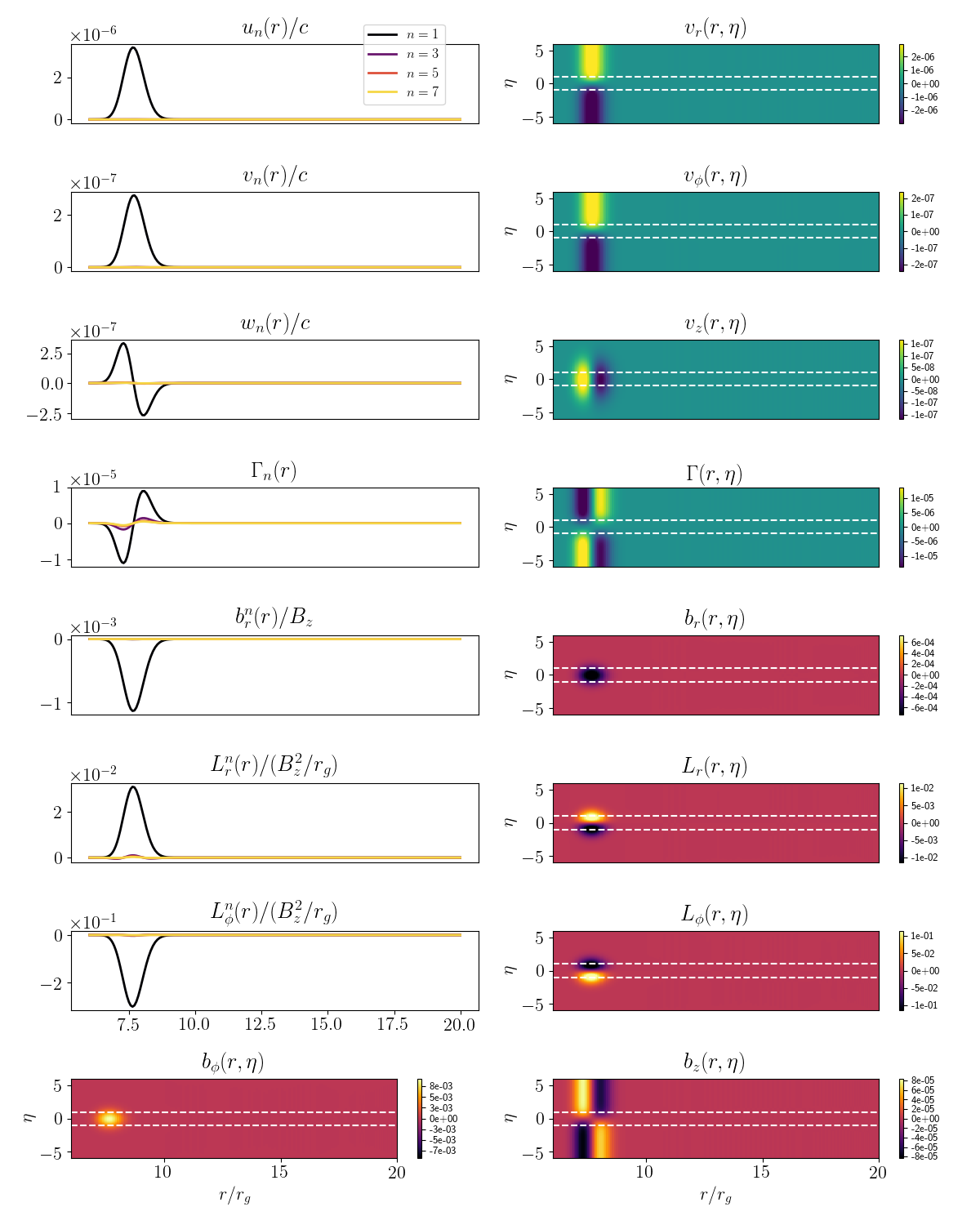}
    \caption{Left: Real parts of the radially variable coefficients
      $[u_n,v_n,w_n,\Gamma_n,b_r^n,L_r^n,L_{\phi}^n](r)$ for the
      radially and vertically fundamental ($l=0$, $k= 1$) trapped
      inertial mode calculated with
      $c_s=0.003c,a=0,r/r_g\in[6,20],N=200,M=7$ and midplane
      $\mathcal{M}_{\textrm{Az}}=0.04$ ($\beta\approx1250$). Right:
      Heatmaps of the fully reconstructed solutions
      $Re[v_r,v_{\phi},v_z,\Gamma,b_r,L_r,L_{\phi}](r,\eta)$. The
      induction equation has been used to find the fields for
      $b_{\phi}(r,\eta)$ (bottom left) and $b_z(r,\eta)$ (bottom
      right). Note that the magnetic field and Lorentz force
      perturbations are proportionately larger in magnitude because they are scaled by much smaller quantities than the velocity perturbations, which are scaled by the speed of light.}\label{cs003zz04a0_cstH_fund}
\end{figure*}

\begin{figure}
    \centering
    \includegraphics[width=.5\textwidth]{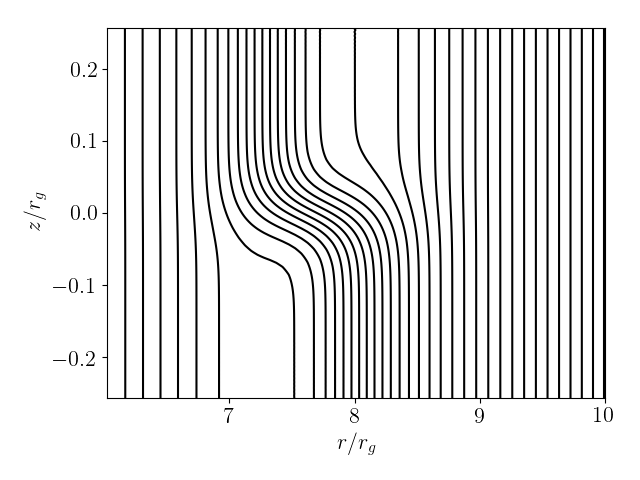}
    \caption{Magnetic field geometry in the $r-z$ plane, close to the trapping region for the $l=0$, $k=1$ r-mode shown in Fig. \ref{cs003zz04a0_cstH_fund}. Magnetic field lines are plotted as evenly spaced contours of the poloidal magnetic flux function. Given the axisymmetry of the field, this can be calculated as $\Psi(r,z)=\int r(\mathcal{A}b_z+B_z)dr,$ where the amplifying factor $\mathcal{A}=1/\max[b_z]$ normalizes the linear perturbation (which is already scaled by $B_z$) so as to make its effect discernible.}\label{fieldz}
\end{figure}

\begin{figure*}
    \centering
    \includegraphics[width=.99\textwidth]{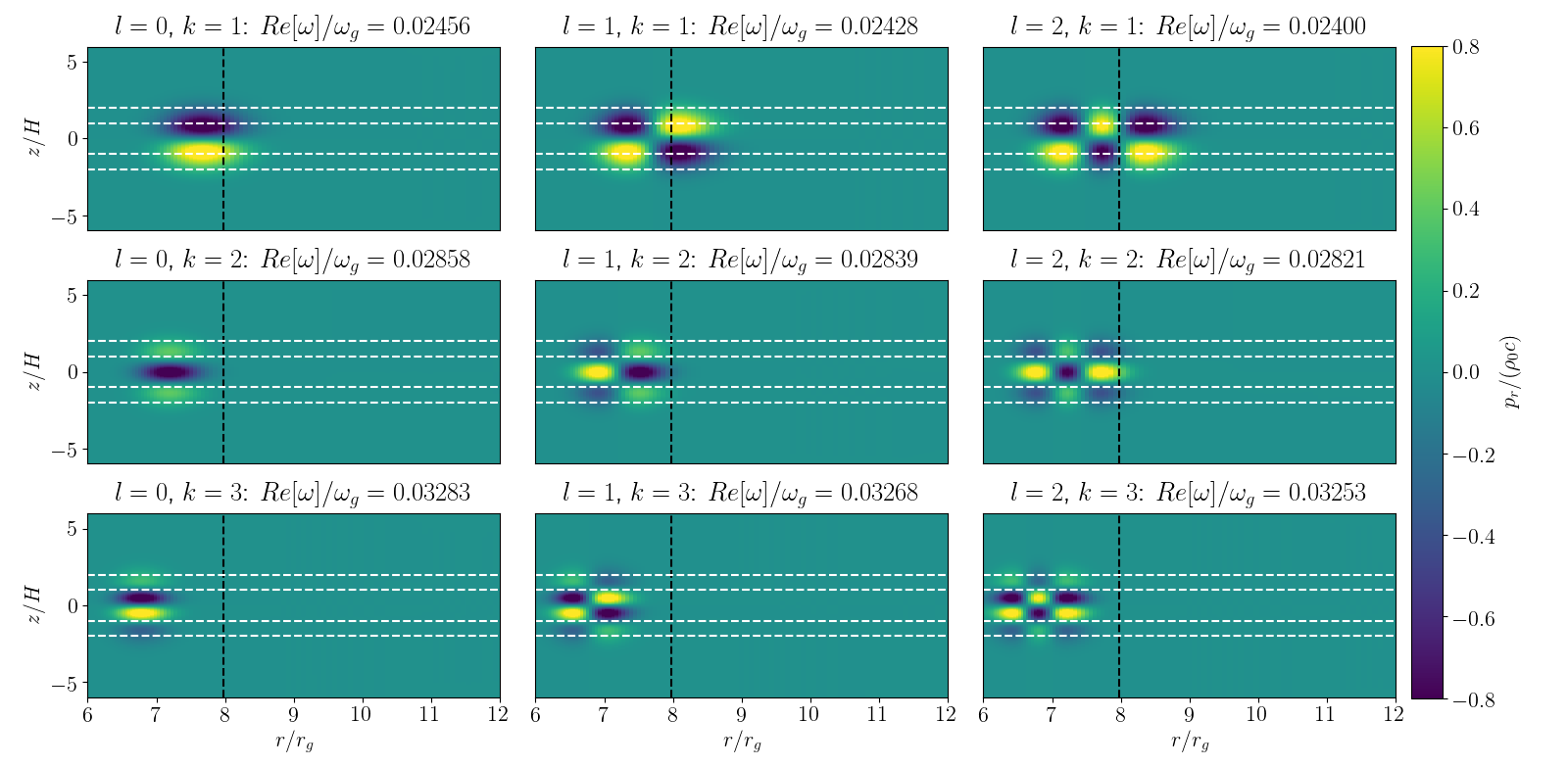}
    \caption{Real parts of the radial momentum perturbation
      $p_r=\rho_0g v_r$ for trapped inertial waves with radial quantum
      numbers $l=0,1,2$ (from left to right) and vertical orders
      $k=1,2,3$ (from top to bottom), calculated as in Fig. \ref{cs003zz04a0_cstH_fund}. The black dashed line indicates $R_{\kappa}$, while the white dashed lines mark $z=\pm H$ and $z=\pm 2H$.}\label{lnvar_cstH}
\end{figure*}

Once the truncated series of coupled eigenvalue problems has been solved, the fully global, ($r,\eta$)-dependent fields can be reconstructed for each perturbation variable $\delta(r,\eta)$ with the summations $\sum_n^M\delta_n(r)\mathcal{B}_n(\eta)$. In the following sections we present fully global r-mode solutions calculated without (\subs \ref{cstH}) and with (\subs \ref{varH}) the coupling provided by radial scale height variation. We use our calculations to find approximations to the critical magnetic field strengths at which r-modes require reflection at the inner boundary for confinement, as well as a correspondence with, and $k_z$ prescription for, the cylindrical model presented in \subs \ref{cylcalc}.

\subsection{Global r-modes in a flat disc}\label{cstH}

In this section we make the simplifying approximation that the scale
height of the disc is purely constant:
$H=c_s/\Omega_z(r_{\textrm{ISCO}})$. This is motivated by the relative
proximity of the trapping region to the inner edge of the disc and
$H$'s weak radial variation within this region. With this approximation
$\del_r\ln H=0$ and $\mathcal{L}_{HH}=0$, so the couplings involving
the coefficients $\lambda_{nm},$ $\nu_{nm}$ and $\mu_{nm}$ drop
out. The absence of a vertical resonance again makes the outer boundary
condition irrelevant, 
since the r-modes should be evanescent at $r_{\textrm{out}}$ as long as it is placed beyond the outer turning point. 

\subsubsection{The global r-mode spectrum}

Solving Equations (\ref{LorEqnsHexp1})-(\ref{LorEqnsHexp7}) with
series truncation at a vertical order $n=M$ produces $M$ sets of
global trapped inertial modes. Each set consists of
modes with similar vertical structure but differing radial structure. As with our cylindrical calculations, we order the members of each set according to the number of radial nodes in $v_r$, e.g.\ $l=0,1,2,..$.
The members of each set are dominated by a single 
component ${\bf{U}}_k$ of order $n=k$, testifying to the weak separability of the problem. As a consequence, all the members of that set express a vertical
profile closely resembling ${\bf{U}}_k$, and for that reason we assign 
a quantum number (or vertical order) `$k$' to the set of modes dominated by the $k$'th order radial coefficients. There are hence two numbers denoting modes:
$l$ and $k$, representing their radial and vertical quantisation respectively.

For example, Fig. \ref{cs003zz04a0_cstH_fund} shows a
representative fundamental r-mode. The $(r,\eta)$ heat maps show the full solution ${\bf{U}}(r,\eta)$, and alongside them we plot the radial profiles of
the various coefficients ${\bf{U}}_n(r)$ that constitute the full
solution. As is clear from the latter, the mode is completely dominated by the
$n=1$ component. We hence assign the vertical quantum number $k=1$ to
this mode, and its radial structure tells us that $l=0$. 
In addition, Fig. \ref{fieldz} shows
the (exaggerated) effect of the mode on the background 
magnetic field lines, close to the trapping region. 

To illustrate the structure of the oscillations within the disc, the component of the linearized momentum perturbation $p_r=\rho_0gv_r$ is shown for similarly calculated modes with $l=0,1,2$ radial and $k=1,2,3$ vertical structures in Fig. \ref{lnvar_cstH}. In the weakly magnetized case with small $\mathcal{M}_{\textrm{Az}},$ the r-modes possessing the same radial node structure but dominated by different $n$ have nearly the same frequencies. With increasing magnetic field strength, however, these groups of solutions separate in both the frequency and physical domains. As foreshadowed by the dependence on $k_z$ observed in our cylindrical calculations, we find that the more vertically complicated modes with larger quantum number $k$ are more strongly affected by the vertical magnetic field. This can be seen in the radial momentum fields shown in Fig. \ref{lnvar_cstH}. The $k=2$ modes (middle row) have been forced further inward by the background magnetic field ($\beta=1250$) than the $k=1$ modes (top row), and the $k=3$ modes (bottom row) even further. Additionally, with more complicated vertical structure, increases in $Re[\omega]$ are larger for modes of a given radial structure. 

As mentioned, the radially and vertically fundamental mode (Fig. \ref{cs003zz04a0_cstH_fund}, top left in Fig. \ref{lnvar_cstH}) is the least likely to be disrupted and most likely to cause an observable luminosity variation, so we focus on its response to the vertical magnetic field. 

\subsubsection{Critical magnetic field strengths}

\begin{figure}
    \centering
    \includegraphics[width=.44\textwidth]{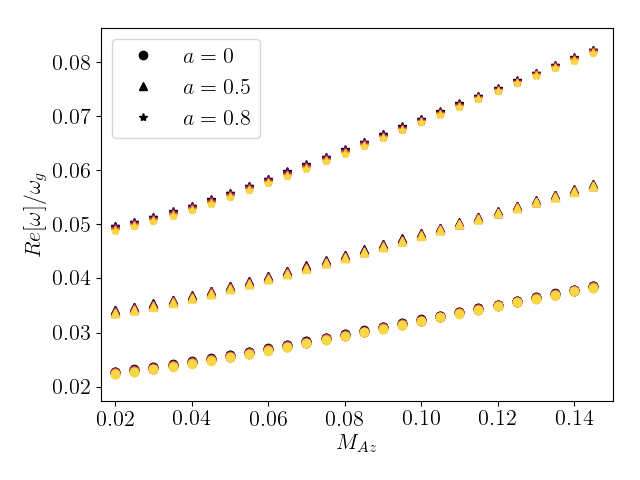}
    \caption{Frequencies for the fundamental $l=0,k=1$ r-mode, calculated using a purely constant scale height (and the boundary condition $\del_rb_r=0$), with varying $c_s$ and midplane $\mathcal{M}_{\textrm{Az}}$, for $a=0$ (circles), $a=0.5$ (triangles) and $a=0.8$ (crosses). Sound speed is indicated with the same colour-map used in Fig. \ref{cs_omDev}, black corresponding to $c_s=0.001c$ and yellow to $c_s=0.01c$ ($r\in[r_{ISCO},r_{ISCO}+14r_g],$ $N=300,$ $M=7$).}\label{cstH_omz}
\end{figure}

\begin{figure}
    \centering
    \includegraphics[width=.5\textwidth]{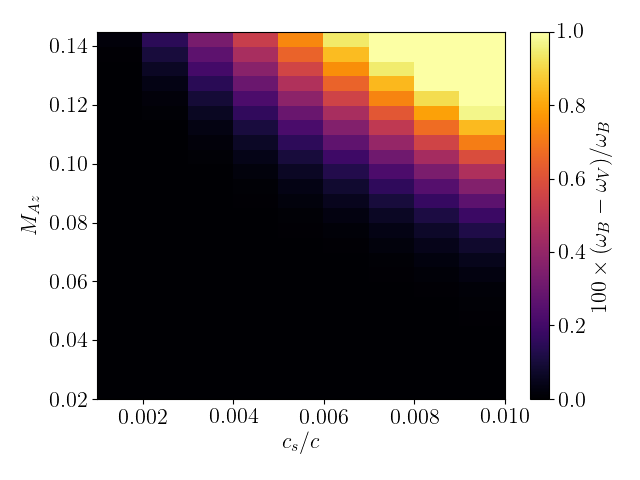}
    \caption{Heatmap of percent differences between frequencies calculated as in Fig. \ref{cstH_omz}, using the boundary conditions $v_r=0$ ($\omega_{\textrm{V}}$) and $\del_rb_r=0,$ ($\omega_{\textrm{B}}$), with varying sound speeds and magnetic field strengths ($a=0$).}\label{cs_omDev}
\end{figure}

We seek a convenient metric for evaluating the critical magnetic field
strength at which trapped inertial mode confinement becomes dependent
on conditions at the inner boundary. At low magnetic
field strengths and sound speeds the frequencies measured
(Fig. \ref{cstH_omz}) are extremely insensitive to the boundary
conditions imposed at $r_{\textrm{in}}$ and $r_{\textrm{out}}$,
agreeing to the precision allowed by our numerical method. However, for sufficiently strong magnetic fields, the frequencies do become
dependent on the inner boundary, with different boundary conditions
producing different frequencies (Fig. \ref{cs_omDev}). 
 We identify the point at which this happens
as one (conservative) measure of the critical magnetic
field strength for r-mode independence from the ISCO. As a point of
comparison, we have also considered the first derivatives of the
radial velocity for modes calculated with rigid boundary conditions.
 The magnetic field strength at which these derivatives becomes
 nonzero
 at the inner boundary provides a very similar, slightly more stringent measure.

\begin{figure}
    \centering
    \includegraphics[width=.5\textwidth]{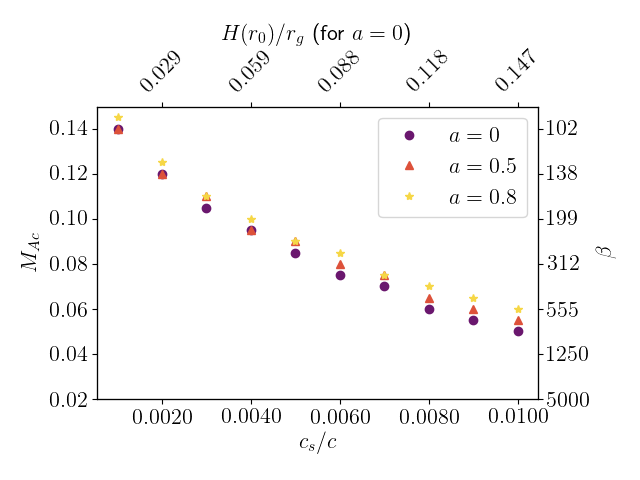}
    \caption{Critical Alfv\'enic Mach number $\mathcal{M}_{\textrm{Ac}}$ found from the constant scale height mode calculations presented in Fig. \ref{cstH_omz} by considering the percent variations in frequency between modes calculated with the inner boundary conditions $v_r=0$ and $\del_rb_r=0$ (see Fig. \ref{cs_omDev})}\label{cstH_zetacrit}
\end{figure}

Fig. \ref{cstH_zetacrit} shows estimates of the critical magnetic
field strengths attained with the former metric. We plot the field
strengths at which the percentage difference between frequencies
calculated with different boundary conditions reaches $0.01\%$ (i.e.,
when $100\times(\omega_{\textrm{V}}-\omega_{\textrm{B}})/\omega_{\textrm{B}}$ reaches
$10^{-2}$, where $\omega_{\textrm{V}}$ and $\omega_{\textrm{B}}$ are
the frequencies calculated with the boundary conditions $v_r=0$ and
$\del_rb_r=0,$ resp.). Fig. \ref{cstH_zetacrit} illustrates the
inverse relationship between disc temperature and r-mode
resilience to vertical magnetic fields. This is to be expected, since trapped inertial modes are less well-confined in hotter, thicker discs even without magnetic fields \citep{FoG08}. For sound speeds ranging from $c_s=0.001c$ to $c_s=0.01c$, we find critical midplane Alfv\'enic Mach numbers of
$\mathcal{M}_{\textrm{A}z}\sim0.14-0.06$ ($\beta\sim100-550$). 
We stress, however, that these values arise from the rather stringent
condition that the frequencies $\omega_{\textrm{V}}$ and $\omega_{\textrm{B}}$ differ by
one-hundredth of a percent, a somewhat arbitrary value. If we allow
frequency discrepancies as large as $0.1\%$ the critical Mach numbers
rises to values above 0.1, and $\beta$ values lower than 100. 
 
The
only other free parameter, the spin of the black hole, has a weaker
effect on the critical field strengths. Larger values of the parameter
$a$ allow r-modes to remain independent of the inner boundary in the
presence of marginally stronger magnetic fields, 
and result in slightly steeper increases in frequency with $\mathcal{M}_{\textrm{A}z}$.

\subsubsection{Cylindrical correspondence}\label{cyl_corr}
In calculating trapped inertial modes in both cylindrical and
stratified disc models, it is natural to ask how solutions found in
the two regimes relate to one another. \citet{Latter15}
showed that local MRI channel modes calculated in the shearing box
correspond to sections in the evanescent regions of much larger,
radially global MRI modes. We consider an analogous relationship
between r-modes in MHD discs with and without vertical density stratification. 

With respect to an appropriate choice of vertical wavenumber,
\citet{FL09} adopted the relationship $k_z\sim \sqrt{\epsilon}/H,$
with $\epsilon$ being of order unity. We find that in general this
relationship holds both with and without the inclusion of radial
variation in $H$. However, we go further, finding a precise value for
the proportionality $\sqrt{\epsilon}$. The rapidity of the convergence
we find with our expansion in the basis functions $F_n$ and $G_n$
suggests the following identification, for $k_z$ in units of $r_g^{-1}$: 
\begin{equation}\label{cylScrip}
k_z
=\dfrac{K_n}{H}.
\end{equation}
In particular we choose $k_z=K_1/H\approx 1.16/H$ to follow our interest in the vertically fundamental, $k=1$ r-mode \citep[cf. Fig. \ref{FnGn}, fig. $1$ in][]{Latter10}. 

\begin{figure}
    \centering
    \includegraphics[width=.5\textwidth]{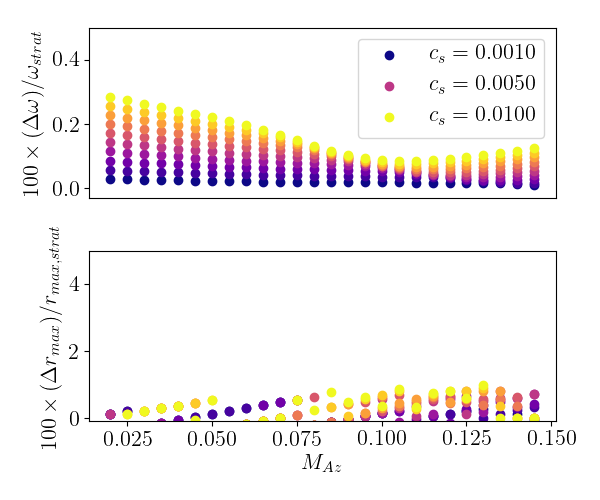}
    \caption{Percent differences between frequencies (top) and radii of maximal $v_r$ (bottom) calculated for fully global r-modes with a purely constant scale height (Fig. \ref{cstH_omz} and Fig. \ref{cstH_zetacrit}) and cylindrical r-modes calculated with the same spin ($a=0$), sound speeds and magnetic field strengths, and $k_z=K_1/H$ (here $\Delta$ indicates the difference between quantities calculated using the two models). Marginally closer relationships were found for larger values of $a$.}\label{K1cyl_corr}
\end{figure}

With this prescription, the frequencies for cylindrical modes closely
follow those of fully global modes. Fig. \ref{K1cyl_corr}(top) shows
the percent difference between frequencies calculated for stratified
r-modes and cylindrical r-modes with $k_z=K_1/H$, at sound speeds
$c_s=0.001-0.01c$, and $a=0$ (similar results are obtained for
$a=0.5,0.8$). In all cases the frequencies differ by less than
$0.5\%$, but the deviation increases from $0.01-0.03\%$ to $0.1-0.3\%$
for larger sound speeds. For comparison, frequencies calculated with
the prescription $k_z=1/H$ give percent differences as large as
$1-10\%$, increasing with larger magnetic field strengths. 
 Fig. \ref{K1cyl_corr}(bottom) shows the percent differences between
 measurements of the radius of maximal $v_r$, $r_{\textrm{max}}$, 
between cylindrical and stratified calculations, which are also less than $1\%$.

The close agreement between frequencies and localisation found through
fully global and unstratified calculations made with $k_z=K_1/H$
suggests that the cylindrical model presented in \subs\ref{cylcalc}
can accurately describe the behaviours of fully global, MHD
r-modes. However, the correct choice of $k_z$ is essential to exploit
this property.

\begin{figure*}
    \centering
    \includegraphics[width=.99\textwidth]{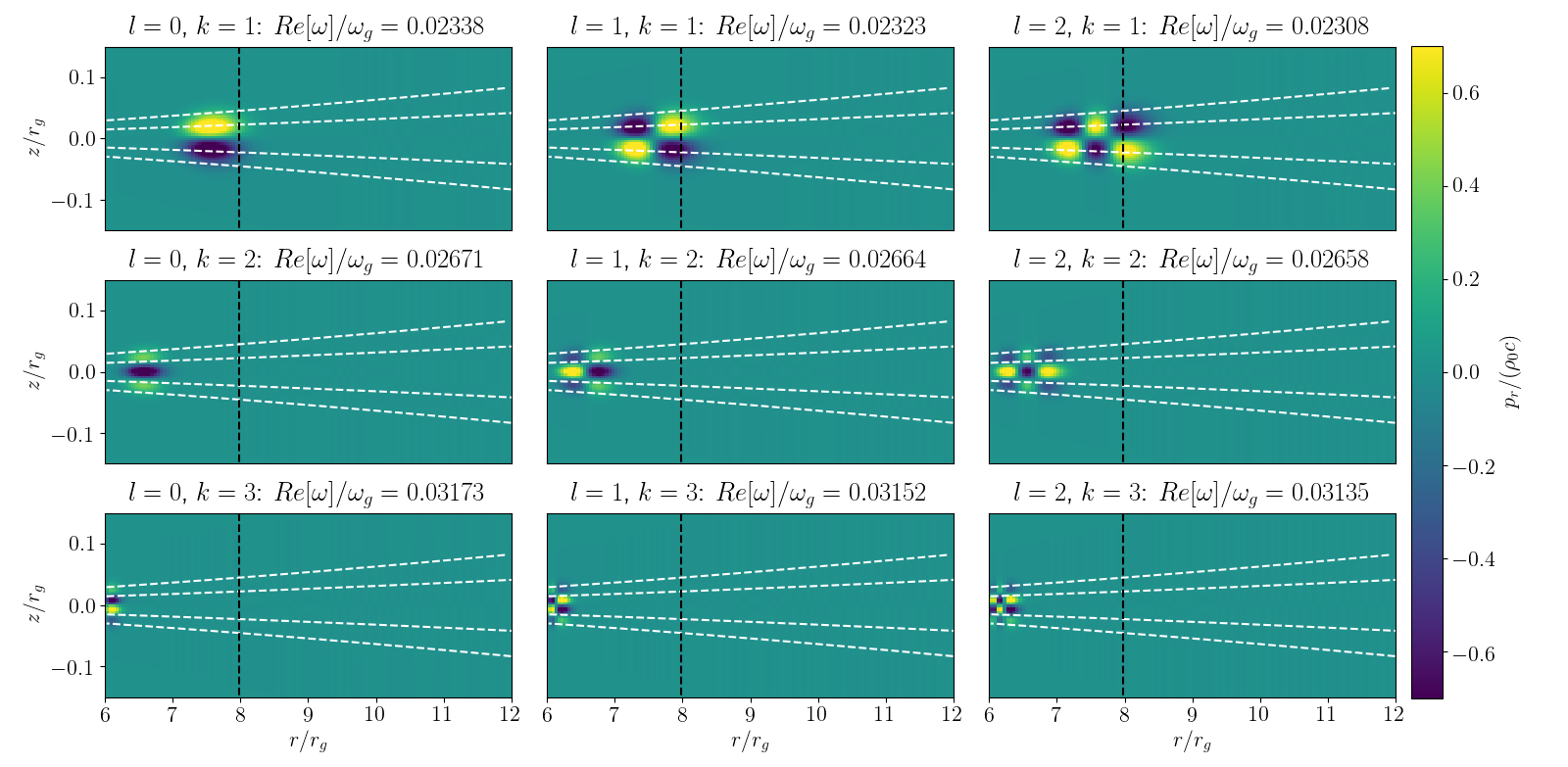}
    \caption{Real parts of the radial momentum perturbation
      $p_r=\rho_0g v_r$ (interpolated from $r,\eta$ to $r,z$
      coordinates) for trapped inertial waves with radial quantum
      numbers $l=0,1,2$ (from left to right) and vertical orders
      $k= 1,2,3$ (from top to bottom), calculated with $c_s=0.001c,a=0,r/r_g\in[6,20],N=300,M=7$, midplane $\mathcal{M}_{\textrm{Az}}=0.04$ ($\beta\approx1250$) and a radially variable scale height. The black dashed line indicates $R_{\kappa}$, while the white dashed lines mark $z=\pm H$ and $z=\pm 2H$.}\label{lnvar_varH}
\end{figure*}

\subsection{The effects of radial scale height variation}\label{varH}

In this section, we include the radial variation of the
isothermal scale height $H=c_s/\Omega_z$, and thus present
fully global, self-consistent calculations of MHD r-modes. We now must contend with  nonzero profiles for $\del_r\ln H$ and $\mathcal{L}_{HH}$, which in turn provide further couplings in Equations (\ref{LorEqnsHexp1})-(\ref{LorEqnsHexp7}) through the coefficients $\nu_{nm}$, $\lambda_{nm}$ and $\mu_{nm}$. The integrals defining these coefficients (see Appendix \ref{Acouple}) are larger than those defining $\theta_{nm},$ $\alpha_{nm}$ and $\gamma_{nm}$, but they couple with the equations only in concert with factors that are $\mathcal{O}(1/r)$ or $\mathcal{O}(1/r^2),$ and we find that their inclusion has little effect on the r-modes. The
largest effect comes from radial variation in the coefficients involving $H^{-1}$, which exacerbates the impact of the background magnetic field. However, the extent of this exacerbation is 
likely exaggerated by our simplifying assumption of a globally constant sound speed.

Introducing radial variation in disc thickness also complicates our choice of boundary conditions, this time at $r_{\textrm{out}}$. In hydrodynamic disc models, a vertical resonance occurs at the radius where $\omega^2=n\Omega_z^2$ \citep{FoG08}, beyond which lies the propagation region for p-modes with frequency $\omega$. \citet{FL09} identified the analogous radius for an unstratified disc threaded by a purely constant vertical field as that where $\omega^2=k_z^2c_s^2$. Given the correspondence found in \subs \ref{cyl_corr}, we place our outer boundary beyond the radius at which $\omega^2=K_1^2\Omega_z^2$ and implement a wave propagation boundary to allow for `tunnelling' of r-modes to the outer regions of the disc. At $r_{\textrm{out}},$ we require $\forall n$ that $\del_ru_n=\textrm{i}k_ru_n,$ with $k_r$ determined using $k_z=K_1/H(r_{\textrm{out}})$ and a frequency calculated for an r-mode using the modified cylindrical model described in \subs\ref{cylcorr_var}. The frequencies calculated for these modes are then complex. However, we find that, as observed by \citet{FoG08} in their hydrodynamic calculations (cf., their Table $1$), the decay rates are negligible for the range of sound speeds considered here.

\subsubsection{Eigenmodes and critical field strengths when $H=H(r)$}

\begin{figure}
    \centering
    \includegraphics[width=.5\textwidth]{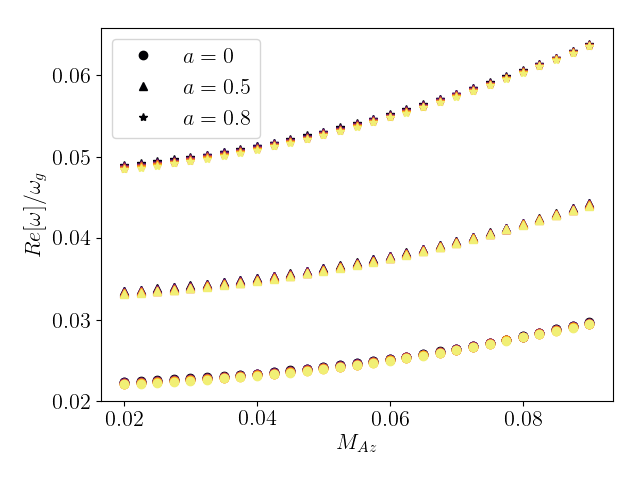}
    \caption{Frequencies for the fully general fundamental r-modes, calculated as in Fig. (\ref{cstH_omz}) but with the inclusion of radial scale height variation, as a function of Alfv\'enic Mach number, sound speed (colour-scale goes from black for $c_s=0.0005c$ to yellow for $c_s=0.004c$) and black-hole spin parameter $a$.}\label{varH_omz}
\end{figure}

The discrete spectrum of global r-modes calculated with the fully self-consistent inclusion of radial scale height variation is similar to that found with the approximation of a purely constant $H$. Truncating our series expansions at vertical order $n=M$ results in $M$ sets of r-modes, each with their own discrete spectrum of radial quantum numbers $l.$ Qualitatively, the response of the r-modes to increasing background magnetic field strength is the same as that observed with a purely constant $H.$ Once again solutions dominated by coefficients of larger vertical order $n=k$ are pushed further towards the ISCO by a given increase in $B_z$ (Fig. \ref{lnvar_varH}), and increases in frequency are comparable (Fig. \ref{varH_omz}). 

\begin{figure}
    \centering
    \includegraphics[width=.5\textwidth]{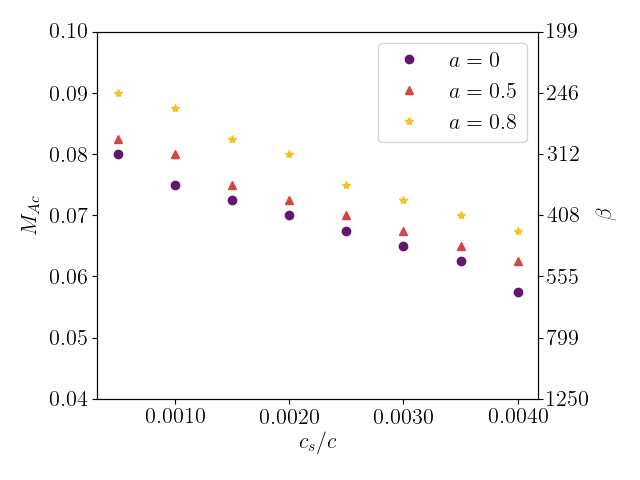}
    \caption{Critical Alfv\'enic Mach number $\mathcal{M}_{\textrm{Ac}}$ found from the variable scale height mode calculations presented in Fig. \ref{varH_omz} by considering the percent variations in frequency between modes calculated with the inner boundary conditions $v_r=0$ and $\del_rb_r=0$.}\label{varH_zetacrit}
\end{figure}

However, the critical field strengths at which frequencies diverge for modes calculated with the boundary conditions $v_r=0$ and $\del_rb_r=0$ at $r_{\textrm{in}}$ are lower by nearly a factor of 2 (Fig. \ref{varH_zetacrit}). This arises through the dependence on disc thickness discussed in \subs \ref{cstH}. For a given sound speed, the r-modes observe a thicker disc at the trapping region with scale height variation than without. Alternatively, one might look to Equation (\ref{bztrap}). For the prescription $k_z\propto H^{-1},$ the contribution to the restoring force provided by the Alfv\'en frequency $k_zV_{\textrm{A}z}$ increases as the mode is forced inward by the vertical magnetic field.

The estimates of critical magnetic field strengths obtained in this section are likely exaggerated by the assumption of global isothermality, though, which results in more rapid flaring of the disc than might be found with more realistic temperature profiles. As mentioned in \subs\ref{model}, however, radial temperature gradients significantly complicate the equilibrium flow, and we do not consider them here. 

\subsubsection{Cylindrical correspondence when $H=H(r)$}\label{cylcorr_var}
Fully global trapped inertial modes calculated with the inclusion of radial scale-height variation do not show a direct correspondence with r-modes calculated in the cylindrical model as presented in \subs\ref{cylcalc}, due to the assumption of a purely constant vertical wavenumber $k_z$ associated with the latter. Suppose now that $k_z=k_z(r)$. For perturbations $\delta$ with the dependence $\delta(r,z)\propto \tilde{\delta}(r)\exp[\textrm{i}k_z(r)z],$ we then have
\begin{equation}\label{dkdr}
    \dfrac{\del \delta}{\del r}
    =e^{\textrm{i}k_zz}\left(
        \dfrac{\textrm{d}\tilde{\delta}}{\textrm{d}r}
        +iz\dfrac{\textrm{d}k_z}{\textrm{d}r}\tilde{\delta}
    \right).
\end{equation}
The dependence on $z$ of the second term on the righthand side eliminates the advantage of separability for which one might assume a plane wave dependence in the first place. However, if in our cylindrical calculations we retain radial variation in using the prescription $k_z=K_1/H(r)$ but assume that the scale height $H$ varies slowly enough for $dk_z/dr$ to be negligible, we find that the coupling of vertical modes is weak enough that the correspondence is still very close, although not as precise as that presented in \subs\ref{cyl_corr}. 

\begin{figure}
    \centering
    \includegraphics[width=.5\textwidth]{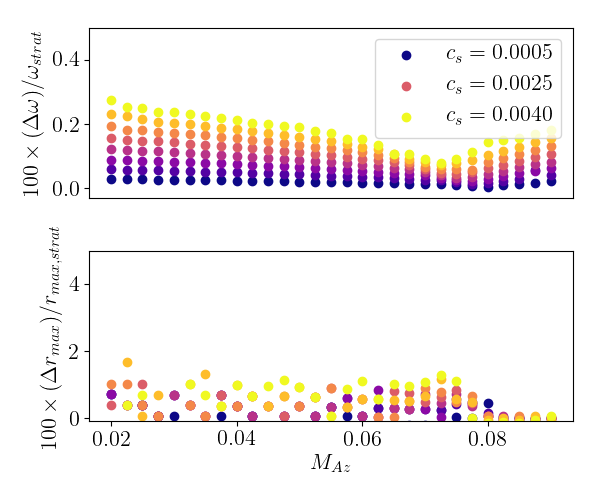}
    \caption{Percent differences between frequencies (top) and radii of maximal $v_r$ (bottom) calculated for fully global r-modes with a radially variable scale height (Fig. \ref{varH_omz} and Fig. \ref{varH_zetacrit}) and cylindrical r-modes calculated with the same spin ($a=0$), sound speeds and magnetic field strengths, and $k_z(r)=K_1/H(r)$ (a similar wave propagation boundary condition to that described in \subs\ref{varH} was used for the cylindrical modes).
    }\label{varH_cyl_corr}
\end{figure}

As in Fig. \ref{K1cyl_corr}, Fig. \ref{varH_cyl_corr} shows the percent difference between frequencies calculated with increasing magnetic field strengths and varying sound speeds $c_s/c=0.0005-0.004$ for stratified modes and cylindrical modes calculated with the $k_z=K_1/H(r)$. Again the percent changes in frequency are less than $1\%$, and percent changes in radii of maximal radial velocity less than $2\%$, but the former increase more rapidly with sound speed than with the approximation of a purely constant scale height. This is perhaps due to the increasing relevance of the radial derivative ignored in Equation (\ref{dkdr}). In terms of the scaled coordinate $\eta,$ this prescription for the vertical wavenumber of a perturbation $\delta(r,\eta)$ might alternatively be written as $\delta(r,\eta)=\tilde{\delta}(r)\exp[\textrm{i}K_1\eta],$ yielding
\begin{equation}
    \dfrac{\del \delta}{\del r}
    =e^{\textrm{i}K_1\eta}
    \left(
        \dfrac{\textrm{d}\tilde{\delta}}{\textrm{d}r}
        -iK_1\eta\dfrac{\textrm{d}\ln H}{\textrm{\textrm{d}}r}\tilde{\delta}
    \right).
\end{equation}
In implementing this prescription for a radially variable $k_z(r)$ in our cylindrical, unstratified model, then, we are omitting the coupling of vertical modes that is included self-consistently in our fully global calculations, and this omission becomes more relevant in hotter, thicker discs. 

\section{Discussion and Conclusions}\label{conc}

We have reconsidered the effects of poloidal and toroidal magnetic fields on trapped inertial waves in magnetised relativistic accretion discs by undertaking cylindrical and fully global linear eigenvalue calculations. Our work expands on and more thoroughly investigates a prediction made by local analyses \citep{FL09}: that a purely constant, vertical magnetic field forces r-modes inward, possibly removing their independence from the inner edge of the disc.

We have verified that poloidal magnetic fields do force r-modes to migrate inwards, and established more realistic estimates for the critical magnetic field strength at which the modes are pushed up against the inner disc edge. However, we find that at lower temperatures MHD inertial modes can remain independent from the inner boundary, even in the presence of relatively strong magnetic fields; the cooler and thinner the disc, the less susceptible the modes are to the fields. We also find that r-mode frequencies can nearly double because of the magnetic field (see Figs. \ref{cstH_omz} and \ref{varH_omz}). This frequency enhancement should be kept in mind when using QPOs to estimate properties of the black-hole-disc system. 

Normal modes calculated with a cylindrical model confirm that an azimuthal field of even equipartition strength has little effect on r-mode frequencies or localisations. Radial gradients in magnetic fields or density, for reasonable power law indices, also have negligible effects. Finally, we show that with an appropriate choice of vertical wavenumber, trapped inertial modes calculated in a simplified cylindrical framework (ideal for global non-linear simulations) closely reproduce the frequencies and localisations of fully global r-modes (to within 1\%).

For an inner accretion disc temperature of $\sim1\textrm{keV}$, isothermal sound speed estimates of
\begin{equation}
c_s=\sqrt{\dfrac{k_BT}{\mu m_p}}
\approx 0.001c
\end{equation}
lead to the survival of the r-mode trapping region in the presence of vertical magnetic fields with midplane Alfv\'enic Mach numbers less than $\mathcal{M}_{\textrm{Az}}\approx 0.08-0.14$ (see Fig. \ref{cstH_zetacrit} and Fig. \ref{varH_zetacrit}). It should be emphasised that these values should be regarded as lower bounds, not only because of the assumption of global isothermality. Moreover, a loss of isolation from the inner boundary need not preclude the existence of trapped inertial waves, given the possibility of wave reflection by steep density gradients or other features. All one can say is that r-modes that are forced up against the ISCO require greater excitation to counteract enhanced energy losses through the inner boundary.

We note that beyond assuming an isothermal equation of state, we have applied other simplifications to our model, in order to more feasibly investigate r-mode propagation in a fully global, MHD context. One such simplification is the omission of a background radial inflow, which intensifies near the ISCO. \citet{FerrThes} considered the effects of radial inflow on trapped inertial modes, and found that decay rates were amplified by the presence of a transonic accretion flow in a viscous disc model. However, the author found that the severity of the damping depends on the location of the sonic point, a location $r_{\textrm{sonic}}<r_{\textrm{ISCO}}$ still allowing r-mode excitation by warps and eccentricities in a hydrodynamic disc. The movement of the trapping region towards the inner boundary by a
poloidal magnetic field would make damping by radial inflow more likely, but not a certainty for sonic points far enough within the ISCO. 

Additionally, although the range of sound speeds considered is motivated by observations, radiation pressure is expected to play a large role in the black hole accretion discs associated with X-ray binaries in the emission states in which HFQPOs are primarily observed. The thickening of the disc associated with this radiation pressure might nullify the increase in critical magnetic field strength we have observed for discs with lower sound speeds, although the inclusion of radiative transfer may modify the dynamics in other ways. 

We have also excluded the impact of radial magnetic fields, which some have suggested might counteract the effects of a purely vertical one \citep{ORetal}, as might disc truncation above and below by a hot corona \citep{Kat17}. Further, we have considered only the effects of large-scale, ordered magnetic fields, for which midplane plasma betas
of $\beta\lesssim500$ are actually rather strong. Although the MRI may saturate near equipartition on small scales, the net flux associated with large-scale poloidal fields such as those considered here strongly affects MHD turbulence and outflows in accretion disc simulations \citep[e.g.,][]{LFO13,BS14,SSAB}. 

Our final take away point is the following. The appearance or not of trapped inertial waves in observations may be determined more by a competition between excitation by disc deformations, and damping by radial inflow and wave leakage, rather than the effect of a large-scale magnetic field on the self-trapping region alone. Models and non-linear simulations including more complicated flow dynamics, thermodynamics, and magnetic field configurations will certainly shed more light on the issue and form the basis for future work.  

\section*{Acknowledgements}
The authors would like to thank the anonymous reviewer for a positive and useful report. J. Dewberry thanks the Cambridge International and Vassar College De Golier Trusts for funding this work.

%%%%%%%%%%%%%%%%%%%%%%%%%%%%%%%%%%%%%%%%%%%%%%%%%%
%%%%%%%%%%%%%%%%% REFERENCES %%%%%%%%%%%%%%%%%%%%%
%\bibliographystyle{mnras}


%%%%%%%%%%%%%%%%%%%%%%%%%%%%%%%%%%%%%%%%%%%%%%%%%%

%%%%%%%%%%%%%%%%% APPENDICES %%%%%%%%%%%%%%%%%%%%%

\appendix

\section{Numerical method for stratified calculations}

\subsection{Coupling coefficients}\label{Acouple}
\begin{figure*}                    
    \includegraphics[width=0.99\textwidth]{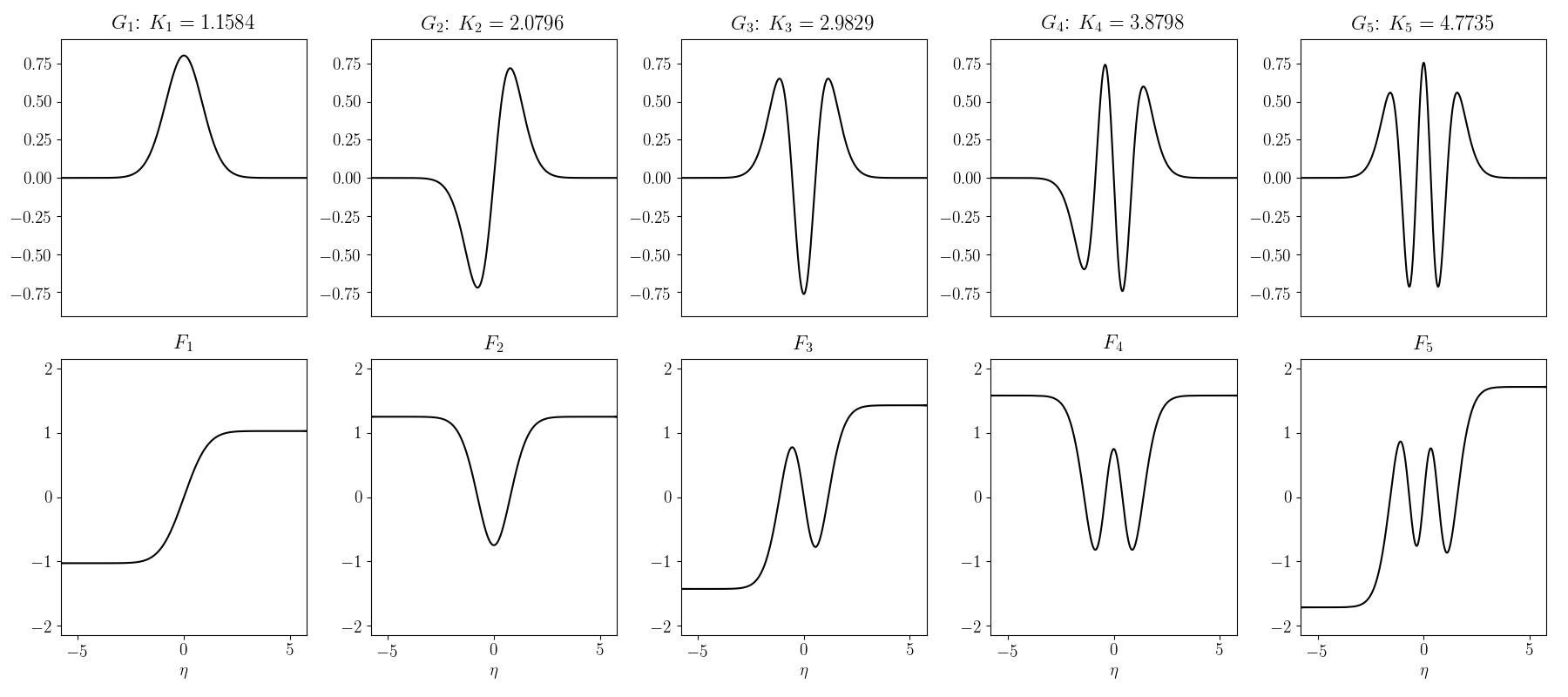}
    \caption{The $n=1,2,3,4,5$ basis functions $G_n$ (top) and $F_n$ (bottom), normalized such that $\int_{-\infty}^{\infty}G_nG_md\eta= \delta_{nm}$ and $\int_{-\infty}^{\infty}gF_nF_md\eta= \delta_{nm}$. The latter are found from the former, which are calculated using pseudo-spectral methods utilizing Whittaker cardinal functions \citep[cf., fig. $1$ in][]{Latter10}. 
    }\label{FnGn}
\end{figure*}

\begin{figure*}
    \centering
    \includegraphics[width=.99\textwidth]{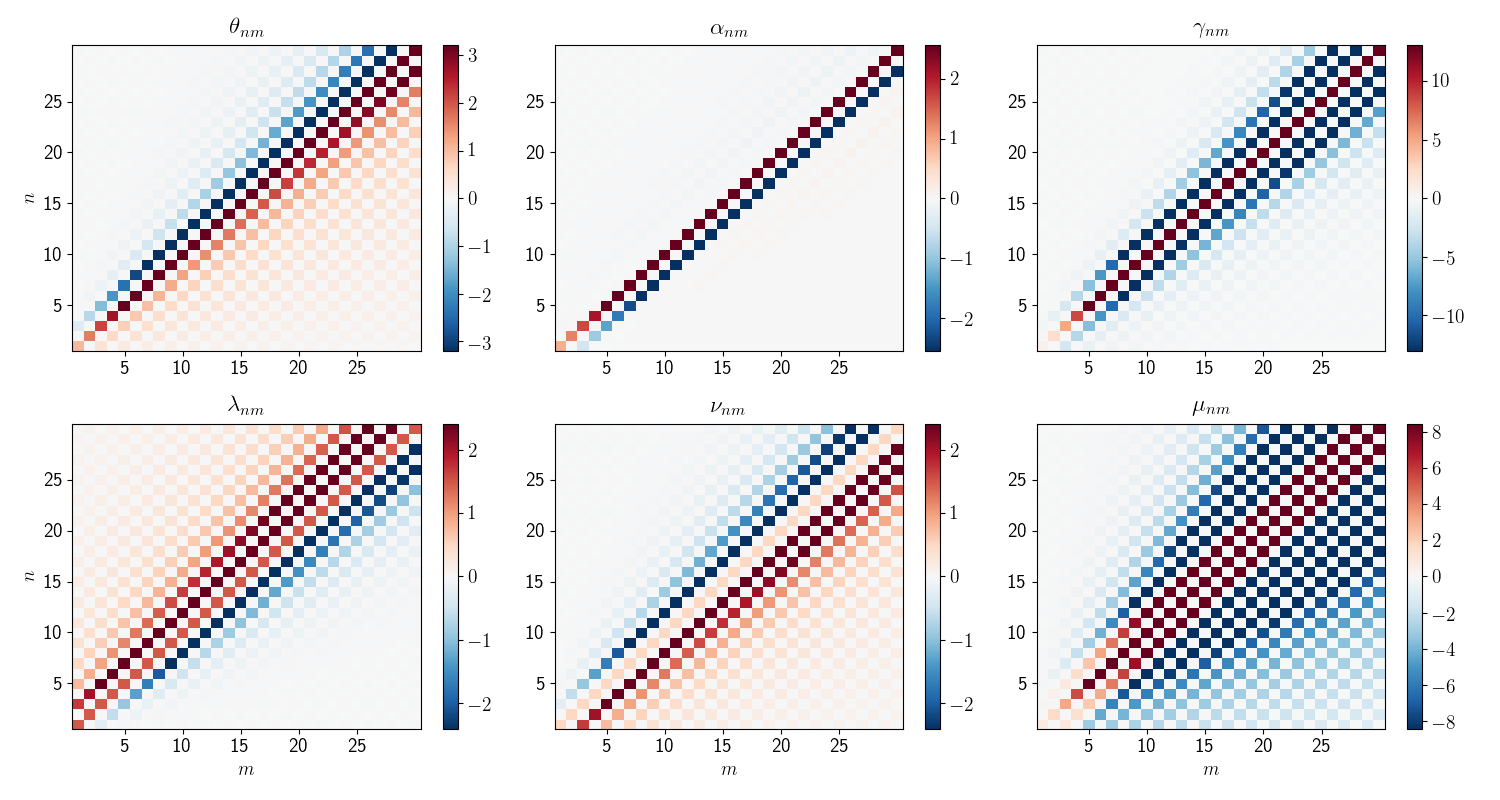}
    \caption{Heatmaps showing the values computed numerically for the coupling integrals $\theta_{nm}$, $\alpha_{nm}$, $\gamma_{nm}$, $\lambda_{nm},$ $\nu_{nm}$ and $\mu_{nm}$. The heatmaps are saturated because the amplitudes of the coupling integrals on the diagonal do not matter (i.e., the equations of a given order $n$ clearly ought to couple very strongly with the equations of order $m=n$).}\label{coeff}
\end{figure*}

Substituting the series expansions discussed in \subs \ref{expBCs} in Equations (\ref{LorEqnsH1})-(\ref{LorEqnsH7}), multiplying by suitable combinations of basis functions $F_n$ and $G_n$ (see Fig. \ref{FnGn}) of arbitrary order and integrating over $\int_{-\infty}^{\infty}d\eta$, the orthogonality relations (\ref{orth1})-(\ref{orth2}) allow for the elimination of many of the infinite sums. This projection onto a given vertical order $n$ is marred only by sums over the other vertical orders $m$ that involve the coupling integrals
\begin{align}
    \theta_{nm}
    &=K_m\int_{-\infty}^{\infty}g^{3/4}F_{n-1}G_md\eta,
\\[10pt]
    \lambda_{nm}
    &=\int_{-\infty}^{\infty} gF_m\left(F_n+\eta K_nG_n\right)d\eta,
\\[10pt]
    \alpha_{nm}
    &=K_n\int_{-\infty}^{\infty}g^{5/4}F_{m-1}G_nd\eta,
\\[10pt]
    \gamma_{nm}
    &=K_m^2\int_{-\infty}^{\infty}g^2F_nF_md\eta,
\\[10pt]
    \nu_{nm}
    &=K_m\int_{-\infty}^{\infty}\eta gF_nG_md\eta,
\\[10pt]
    \mu_{nm}
    &=K_m^2\int_{-\infty}^{\infty}\eta^2 g^2F_nF_md\eta.
\end{align}
These integrals are evaluated by first calculating the basis functions $F_n,G_n$ and eigenvalues $K_n$, then integrating numerically over the range of $\eta$ within which the normalized integrands are $>10^{-4}.$ The numerical values for the integrals computed with $n,m\in[1,30]$ are shown in the heatmaps in Fig. (\ref{coeff}). For small $n,$ the integrals go to zero very quickly with increasing $m,$ such that the couplings are very weak for equations with very different vertical order. Additionally, the heatmaps in Fig. (\ref{coeff}) illustrate the symmetry in Equations (\ref{LorEqnsH1})-(\ref{LorEqnsH7}) ensuring that mode couplings are restricted to either even or odd parity.

\subsection{Numerical convergence}\label{converge}
To check for convergence with increasing series truncation order $M$, we track both the normal mode frequencies and the relative energetic contributions from each ${\bf{U}}_n$ to the full solution ${\bf{U}}.$ For $|f|^2=ff^*$, we quantify the latter through norms defined for each variable, at each vertical order $n$, as 

\begin{align}
    v_r:&\qquad\int_{-\infty}^{\infty}\int_{r_{in}}^{r_{\textrm{out}}}\rho_0g|u_nF_n|^2drd\eta,
\\[10pt]
    v_{\phi}:&\qquad\int_{-\infty}^{\infty}\int_{r_{in}}^{r_{\textrm{out}}}\rho_0g|v_nF_n|^2drd\eta,
\\[10pt]
    v_z:&\qquad\int_{-\infty}^{\infty}\int_{r_{in}}^{r_{\textrm{out}}}\rho_0g|w_ng^{1/4}F_{n-1}|^2drd\eta,
\\[10pt]
    \Gamma:&\qquad\int_{-\infty}^{\infty}\int_{r_{in}}^{r_{\textrm{out}}}|c_s^2\Gamma_nF_n|^2drd\eta,
\\[10pt]
    b_r:&\qquad\int_{-\infty}^{\infty}\int_{r_{in}}^{r_{\textrm{out}}}|b_r^nG_n|^2drd\eta,
\\[10pt]
    L_r:&\qquad\int_{-\infty}^{\infty}\int_{r_{in}}^{r_{\textrm{out}}}|L_r^ngF_n|^2drd\eta,
\\[10pt]
    L_{\phi}:&\qquad\int_{-\infty}^{\infty}\int_{r_{in}}^{r_{\textrm{out}}}|L_{\phi}^ngF_n|^2drd\eta.
\end{align}

With the expansions discussed in \subs \ref{expBCs}, the energetic contributions from the non-dominant $n$ are at least two orders of magnitude smaller than that from the dominant $n$, and quickly fall off with increasing series truncation to values smaller than the error introduced by our Chebyshev collocation method. Beyond truncation at $M=5$, changes in frequency for the fundamental $k=1$ r-mode are less than $10^{-4}\%$, and continue to fall with increasing $M$. Frequencies found with increasing $M$ and $\mathcal{M}_{\textrm{A}z}$ are shown for in Tables \ref{Mconv_cstH} and \ref{Mconv_varH} for constant and variable scale height calculations (resp.). Like Table \ref{Mconv_varH}, Table \ref{Mconv_varH_noHcoup} shows frequencies calculated with variable $H(r)$ but with the coupling provided by the radial derivatives of $H$ excluded. 

It should be noted that we have more trouble resolving the coefficients ${\bf{U}}_n$ and frequencies at lower magnetic field strengths ($\mathcal{\textrm{M}}_{\textrm{A}z}\lesssim0.04$), when larger sound speeds ($c_s\gtrsim0.007c$ for constant $H$ and $c_s\gtrsim0.003c$ for variable $H$) are used. Beyond the largest values of $c_s$ considered in \subs\ref{cstH} and \subs\ref{varH} ($0.01c$ and $0.004c$, resp.), these resolution issues extend throughout the full range of magnetic field strengths considered. This indicates that the series expansions given in \subs\ref{expBCs} are less appropriate in thicker discs, and highlights the importance of disc thinness to r-mode trapping in magnetized contexts.

\begin{table*}
	\centering
	\begin{tabular}{lccccccr} % six columns, alignment for each
		\hline
		$\mathcal{M}_{\textrm{A}z}$ & 0.02 & 0.04 & 0.06 & 0.08 &  0.1 & 0.12 & 0.14\\
		\hline
		$\sim\beta$ & 5000 & 1250 & 556 & 312 &  200 & 139 & 102\\
		\hline
		$M=1$ & 0.022789  & 0.024658 & 0.02708  & 0.029722 & 0.032439 & 0.035166 & 0.037882\\
        $M=3$ & 0.022791  & 0.02466  & 0.027082 & 0.029726 & 0.032443 & 0.03517  & 0.037886\\
        $M=5$ & 0.022791  & 0.024661 & 0.027083 & 0.029726 & 0.032443 & 0.035171 & 0.037887\\
        $M=7$ & 0.022791  & 0.024661 & 0.027083 & 0.029727 & 0.032444 & 0.035172 & 0.037887\\
        $M=9$ & 0.022791  & 0.024661 & 0.027083 & 0.029727 & 0.032444 & 0.035172 & 0.037887\\
        $M=11$ & 0.022791 & 0.024661 & 0.027084 & 0.029727 & 0.032444 & 0.035172 & 0.037887\\
        $M=13$ & 0.022791 & 0.024661 & 0.027084 & 0.029727 & 0.032444 & 0.035173 & 0.037888\\
        $M=15$ & 0.022791 & 0.024661 & 0.027084 & 0.029727 & 0.032444 & 0.035173 & 0.037888\\
		\hline
	\end{tabular}
	\caption{ Frequencies $Re[\omega]/\omega_g$ for the fundamental $l=0$, $k= 1$ trapped inertial mode, calculated with a purely constant scale height and the boundary condition $\del_rb_r=0$, for varying magnetic field strength and series truncation order $M$ ($a=0,c_s=0.001,r/r_g\in[6,20],N=200$). }\label{Mconv_cstH}
\end{table*}

\begin{table*}
	\centering
	\begin{tabular}{lcccccccr} % six columns, alignment for each
		\hline
		$\mathcal{M}_{\textrm{A}z}$ & 0.02  &  0.03 & 0.04 & 0.05 & 0.06 &  0.07 & 0.08 & 0.09\\
		\hline
		$\sim\beta$ & 5000 & 2222 & 1250 & 800 & 555 & 408 & 312 & 246\\
		\hline
		$M=1$ & 0.022342 & 0.022768 & 0.023375 & 0.024174 & 0.025184 & 0.026431 & 0.027975 & 0.029662\\
        $M=3$ & 0.022346 & 0.022772 & 0.023379 & 0.024178 & 0.025187 & 0.026433 & 0.027977 & 0.029667\\
        $M=5$ & 0.022347 & 0.022773 & 0.023379 & 0.024178 & 0.025188 & 0.026434 & 0.027977 & 0.029668\\
        $M=7$ & 0.022347 & 0.022773 & 0.023379 & 0.024179 & 0.025188 & 0.026434 & 0.027977 & 0.029669\\
        $M=9$ & 0.022347 & 0.022773 & 0.02338  & 0.024179 & 0.025188 & 0.026434 & 0.027978 & 0.029669\\
        $M=11$ & 0.022347 & 0.022773 & 0.02338  & 0.024179 & 0.025188 & 0.026434 & 0.027978 & 0.029669\\
        $M=13$ & 0.022347 & 0.022773 & 0.02338  & 0.024179 & 0.025188 & 0.026434 & 0.027978 & 0.02967\\
        $M=15$ & 0.022347 & 0.022773 & 0.02338  & 0.024179 & 0.025188 & 0.026435 & 0.027978 & 0.02967\\
		\hline
	\end{tabular}
	\caption{Frequencies $Re[\omega]/\omega_g$for the fundamental $l=0$, $k=1$ trapped inertial mode, calculated with a variable scale height and the boundary condition $\del_rb_r=0$, for varying magnetic field strength and series truncation order $M$ ($a=0,c_s=0.001,r/r_g\in[6,20],N=200$).}\label{Mconv_varH}
\end{table*}

\begin{table*}
	\centering
	\begin{tabular}{lcccccccr} % six columns, alignment for each
		\hline
		$\mathcal{M}_{\textrm{A}z}$ & 0.02  &  0.03 & 0.04 & 0.05 & 0.06 &  0.07 & 0.08 & 0.09\\
		\hline
		$\sim\beta$ & 5000 & 2222 & 1250 & 800 & 555 & 408 & 312 & 246\\
		\hline
		$M=1$ & 0.022342 & 0.022768 & 0.023375 & 0.024175 & 0.025184 & 0.026431 & 0.027977 & 0.029665\\
        $M=3$ & 0.022346 & 0.022772 & 0.023379 & 0.024178 & 0.025188 & 0.026434 & 0.027978 & 0.029671\\
        $M=5$ & 0.022347 & 0.022773 & 0.02338  & 0.024179 & 0.025188 & 0.026435 & 0.027979 & 0.029672\\
        $M=7$ & 0.022347 & 0.022773 & 0.02338  & 0.024179 & 0.025188 & 0.026435 & 0.027979 & 0.029672\\
        $M=9$ & 0.022347 & 0.022773 & 0.02338  & 0.024179 & 0.025189 & 0.026435 & 0.027979 & 0.029673\\
        $M=11$& 0.022347 & 0.022773 & 0.02338  & 0.024179 & 0.025189 & 0.026435 & 0.027979 & 0.029673\\
        $M=13$& 0.022347 & 0.022773 & 0.02338  & 0.024179 & 0.025189 & 0.026435 & 0.027979 & 0.029673\\
        $M=15$& 0.022347 & 0.022773 & 0.02338  & 0.024179 & 0.025189 & 0.026435 & 0.027979 & 0.029673\\
		\hline
	\end{tabular}
	\caption{Frequencies $Re[\omega]/\omega_g$ for the fundamental $l=0$, $k=1$ trapped inertial mode, calculated as in Table \ref{Mconv_varH}, but with the coupling provided by the radial derivatives of $H$ excluded.}\label{Mconv_varH_noHcoup}
\end{table*}

%%%%%%%%%%%%%%%%%%%%%%%%%%%%%%%%%%%%%%%%%%%%%%%%%%

% Don't change these lines
\bsp	% typesetting comment
\label{lastpage}
\end{document}